\documentclass[preprint,tightenlines]{revtex4-1}  
\usepackage[]{graphicx}
\usepackage{color}
\usepackage{bm}
\usepackage{amsmath}
\usepackage{amssymb}
\usepackage{url}
\usepackage[makeroom]{cancel}
\usepackage{changes}
\usepackage{enumitem}
\definecolor{purple}{cmyk}{0.1,0.9,0,0.1}
\definecolor{dgreen}{cmyk}{0.8,0,0.8,0.2}
\definechangesauthor[name={Vitaly}, color=dgreen]{vr}
\definechangesauthor[name={Sophya}, color=purple]{sg}
\definechangesauthor[name={Julian}, color=blue]{js}
\newcommand{\be}{\begin{equation}}
\newcommand{\ee}{\end{equation}} 
\newcommand{\bea}{\begin{eqnarray}}
\newcommand{\eea}{\end{eqnarray}} 
\newcommand{\bra}{\langle}
\newcommand{\ket}{\rangle} 
\newcommand{\grad}{\nabla} 
\newcommand{\pd}{\partial} 
\newcommand{\ba}{\begin{array}}
\newcommand{\ea}{\end{array}}

\begin{document}

\title
{
  Factorized electron-nuclear dynamics with effective complex potential: on-the-fly implementation for  H$_2^+$ in a laser field
}

\author{Julian Stetzler}
\author{Sophya Garashchuk}\email{garashchuk@sc.edu}
\author{Vitaly Rassolov}\email{rassolov@mailbox.sc.edu}
\affiliation{Department of Chemistry \& Biochemistry, University of South Carolina, Columbia, South Carolina 29208}
\date{\today}

\begin{abstract}
 Conventional theoretical and computational approaches to 
 fully coupled quantum molecular dynamics, i.e. when both the electrons and nuclei  are  treated as quantum-mechanical particles,  are  impractical  for all but the smallest chemical systems.  In this paper we describe the formalism and  implementation of  the  Factorized Electron Nuclear Dynamic (FENDy) with effective complex potential [J. Chem. Theory Comput. {\bf 19} (2023), pp 1393-1408],  which goes beyond the established  framework of  the Born-Oppenheimer approximation or Born-Huang expansion of the molecular wavefunction.   This method is based on the exact factorization of the  molecular wavefunction, with the nuclei evolving under a complex time-dependent potential which captures the key features of dynamics in the nuclear subspace. The complementary  electronic  component of the molecular wavefunction is normalized to one for all nuclear configurations.    We implement and employ FENDy to model the dynamics of  H$_2^+$ molecular ion under  a femtosecond laser pulse. The electronic wavefunction is represented within the  standard electronic structure bases without referencing the electronic eigenstates. 
  The nuclear wavefunction is represented as a quantum-trajectory  ensemble, which  in principle circumvents  the exponential scaling of the numerical cost with the system size. 
   The challenging evaluation of the gradients on  unstructured grids  is  performed  by projection on  auxiliary bases.
   
   \vspace{.5in}
\end{abstract}

\maketitle
\newpage

\section{Introduction}
  One of '33 Unresolved Questions in Nanoscience and Nanotechnology'  posed by  the  authors  in a recent  focus article in  ACS Nano \cite{mirkin2025-nano}  is  "How can we model materials across length scales where classical or quantum mechanics alone are  inadequate to match theory and experiment?" 
Such  quantum/classical delineation is typically understood  as 'capturing electronic
 properties of small systems' with quantum mechanics,   while   'modeling larger systems using
 empirical or quantum-derived forces' with classical molecular dynamics, i.e. trajectories, and Monte Carlo methods \cite{mirkin2025-nano}.  While both frameworks provide fundamental  insights into the properties and behavior of molecular systems,  
  current experiments and theories,  e.g.  those focused on  coherent and   strongly   coupled quantum motion of the nuclei,  plasmons  and light \cite{delor2023,coherence2024,PolaritonReview,reichman2025},  
   go  well beyond the above-mentioned quantum/classical separation of time- and length-scales associated with electrons and nuclei or atoms, respectively.  
  However,  there is no established theoretical framework which readily incorporates   the nuclear quantum effects (NQE) into the trajectory dynamics   of  large molecular systems,   combined with the  spatially 'localized'  description of  electronic wavefunctions  as wavepackets  consisting of  hundreds and thousands of electronic eigenstates.   
   Some recent theoretical studies involving hundreds of electronic states include, for example, behavior plasmonic silver–platinum nanoparticles in photocatalysts \cite{schatz2025}, nonradiative exciton dynamics in monolayer black phosphorus \cite{akimov25-50states}, and  soft x-ray near-edge spectra in transition metal complexes \cite{pak2025}, while simulation of plasmons in nanorods of up to 2160 silver atoms extend to 24,000 electrons\cite{jakowski2025}.  
   When it comes to the quantum behavior of the nuclei, the NQEs are most often associated with motion of light particles, such as protons and their isotopologues \cite{markland2018},  though quantum tunneling of heavy atoms  is gaining recognition as well (e.g.   the heavy atom tunneling associated  with  the ring expansion of fluorenylazirines \cite{sander2023}).   
   Recent state-of-the-art rigorous quantum dynamics studies (implemented in up to 12 dimensions)  based on stationary basis or discrete variable representations \cite{dvr-chapter} of nuclear wavefunctions and sophisticated contraction schemes, helped  explain experiments involving  molecules and clusters  in gas phase and in confinement \cite{bowman2025,carrington2025,bacic2020,bacic2023,bacic2025}. 
   However, given the exponential scaling  of  exact methods with the system size and the need to include NQEs  and  multiple electronic states, 
   motivates theoretical development departing from the basis description of the nuclear wavefunctions and the   Born-Huang representation of the molecular wavefunction \cite{TullyMetalsReview}. 
   
One such development   is  the exact factorization (XF) method, introduced into   theoretical chemistry  by Abedi, Maitra and Gross in  2010 \cite{amg10}. The XF   is  an approach   to   quantum molecular dynamics based on the product form of time-dependent multi-component wavefunctions, which for clarity of discussion here  will be taken  as the electron-nuclear wavefunction,  
    $$  \Psi(\bm R,q,t) = \underbrace{\psi(q,t)}_{nuclear}\underbrace{ \Phi(\bm R, q,t)}_{electronic}, $$
   subject to the partial normalization condition  of $\Phi$ in the electronic space. Here,  vector  $\bm R$ represents electronic coordinates. The electronic wavefunction is normalized to $1$,   $\bra\Phi|\Phi\ket_{\bm R}=1$,  for any configuration of the nuclei, denoted here as a single coordinate  $q$,  at all times, $t$.  For such factorization  to be 'exact',  the electronic component, $\Phi$,  has to be  a function of  both the electronic and nuclear coordinates. The  XF  representation can be viewed as  a  way of  dealing  with large numbers of  electronic states, by  $\Phi(\bm R,q,t)$ being an electronic wavepacket, and as a starting point for well-defined approximations to the  dynamics of both wavefunction components, necessary for applications to large molecular systems. 
 
Recent theoretical developments in XF include extensions to polaritonic chemistry \cite{NeepaPolariton,TokatlyPolariton,XFPhotonNuclear}, a new XF-based theory of electronic friction \cite{RoccoIreneFriction}, and the factorized electron-nuclear dynamics (FENDy) with a complex potential \cite{Fendy}. One major application of XF ideas was the development of the Coupled Trajectory-Mixed Quantum Classical (CT-MQC) method \cite{CTMQC}. Further applications include decoherence corrections derived from the XF to augment surface hopping. These methods were recently added to the Libra software package for use with both model and atomistic systems \cite{XFLibra}.

To the best of our knowledge, the largest  chemical application to date  is the study of the ring-opening of oxirane, where the approximate CT-MQC method was applied to a realistic molecular system using on-the-fly Density Functional Theory computation of the electronic structure \cite{CTMQCApplied,XFOxirane,IbeleNamdReview,akimov24}. However,   when it comes to use of   XF as an  exact method  applications are limited to  small model systems, such as Tully models \cite{XFTullyModels,Fendy}, one-dimensional H$_2^+$ in a laser field with soft coulomb interactions \cite{ElectronLocalisation,XFH2LaserField}, and the Shin-Metiu \cite{sm95,ShinMetiuXF} model where accurate solutions on the full-space can be used to construct the XF wavefunctions.  In these studies, XF provided conceptual understanding of molecular Berry phases \cite{XFBerryPhase},  and the forces driving nuclear dynamics in non-adiabatic systems \cite{BasileAgostiniSteps,XFSteps}.   The reason is  that the numerical solution of the coupled  electronic and nuclear Schr\"odinger  equations  within the  XF, presents major challenges  associated the  singularity of the non-classical  momentum, $r(q,t)$,   
   $$ r(q,t):=\frac{\bm \grad_{q}|\psi(q,t)|}{|\psi(q,t)|}, $$   in the regime of  diverging XF nuclear probability density, $|\psi(q,t)|^2$, and in the low-amplitude regions of  $\psi(q,t)$, the latter due to  the partial normalization condition of  the electronic wavefunction $\Phi(\bm R, q, t)$  imposed at {\it all} $q$. 
 The origin  and handling of the singularity in $r(q,t)$,  including the use of masking functions as a way of   reducing   the associated noise \cite{neepanumericalxf} 
   were recently discussed  in the literature  \cite{lorinxf,stetzler2025}.  

  In this paper we  report the first implementation of  the  factorized electron-nuclear dynamics with complex potential  in the nuclear subspace  for the H$_2^+$ molecule, which has been widely studied as a model system for electron-nuclear dynamics with photoexcitations. \cite{Dunn1968,Argyros1974,Maas1975,Ozenne1976,Saha1980,Cehlkowski1995,Giusti1995,Kawata1999a,Kawata1999b,Lebedev2003,Kreibich04,Kono2004}. We employ the basis description of an electron (3 DOFs)  interacting  with two protons via the Coulomb potential. The   nuclear XF wavefunction is represented by   the  quantum trajectory  ensemble -- an approach which is  potentially scalable to large systems   \cite{wyatt_book} --  and  the  system is subjected to  a laser pulse leading to  photodissociation of H$_2^+$.     
   
   In the remainder of this paper,  Section \ref{sec:theory}  presents the model and theoretical background. Details of implementation are described in  Section \ref{sec:Implementation};  the results and discussion are given in Section \ref{sec:results}; Section \ref{sec:summary} gives a summary and outlook.

\section{Theory} \label{sec:theory}
For  clarity  we will first introduce the  molecular model, and then present the theoretical approach using notations specific to the model.  Cursive bold letters denote vectors, while capital bold letters are reserved for the matrices. Generalization of the theory to multidimensional nuclear subspace can be found in Refs \cite{amg10,Fendy}. Integration over the electronic subspace is denoted $\bra ... \ket_{\bm{R}}$.
\subsection{The Model}
We will consider dynamics of H$_2^+$  --  a system comprised of one  electron and two protons  -- subjected  to a laser field. The system is described in the internal molecular coordinate $q$, that describes the bond length, with no rotation.
With that the total Hamiltonian (in atomic units) is
\be \hat{H}=-\frac{1}{2M}\nabla^2_{q}+\hat{H}^{BO}+\hat{V}^{ext}({\bm R},q,t),  \ee
where $\hat{H}^{BO}$ is the usual Born-Oppenheimer electronic hamiltonian, 
\be  \hat{H}^{BO}:=-\frac{1}{2}\nabla^2_{\bm R}-\frac{1}{\sqrt{x^2+y^2+(z+q/2)^2}}-\frac{1}{\sqrt{x^2+y^2+(z-q/2)^2}}+\frac{1}{q}. \label{eq:HBO}\ee 
The vector  ${\bm R=(x,y,z)}$ denotes the electrons in full three-dimensional Cartesian coordinates, and $M$ is the reduced mass of the nuclei. The external potential, $\hat{V}^{ext}({\bm R},q,t)$, is  the  laser field in the length gauge. It is defined as the scalar product of 
 the  electric field vector, $\bm E(t)$,  and the electronic coordinate  $\bm R$. The electric field is taken   as  a monochromatic beam in a Gaussian envelope along the electronic $z$-direction (along the bond axis), 
\be {\bm E}(t)=\left(0,0,\varepsilon e^{-\alpha (t-t_0)^2}\sin(\omega t)\right), \ee 
yielding  \be \hat{V}^{ext}({\bm R},q,t)={\bm E}(t)\cdot\bm R = \varepsilon  z e^{-\alpha (t-t_0)^2}\sin(\omega t). \ee
 The numerical values of the parameters,  given in Table \ref{tab:ModelParams},  correspond to a femtosecond (approx 40 a.u.) laser pulse with a wavelength of 106 nm and a peak intensity on the order of 10$^{14}$ W/cm$^2$. The length of the pulse is defined by the width of the gaussian envelope $\alpha$ and the wavelength by the frequency $\omega$. The parameter $\varepsilon$ defines the peak intensity and $t_0$ defines the time of peak intensity. This model allows us  to examine  the effect of the  electron-nuclear coupling  on dynamics directly by working without  approximations  to  the electron correlation necessary for multi-electron systems, while the electron is described within standard basis sets of quantum chemistry. 

 \begin{table}
      \centering
      \begin{tabular}{c|c|c|c|c}\hline
      \multicolumn{5}{c}{Model parameters (a.u.)} \\
      \hline
    $M=918.076336$ & $\varepsilon=0.035$ & $t_0=40$ & $\alpha=0.01$ & $\omega=0.420604544$ \\\hline
      \end{tabular}
      \caption{\footnotesize The model parameters used for the numeric analysis. $M$ is the reduced mass of the two protons. The remaining parameters define the laser pulse. The frequency of the laser is $\omega$ and corresponds to a wavelength of 106 nm; its peak intensity is $\varepsilon$ giving an intensity on the order 10$^{14}$ W/cm$^2$ in SI units occurring at time $t_0$. The envelope width $\alpha$ is about 40 a.u.}
      \label{tab:ModelParams}
\end{table}   

\subsection{The exact factorization approach with complex nuclear  potential}\label{sec:FENDy}
In this section we outline  the  exact factorization (XF)  formalism, specifically  FENDy  presented in Ref. \cite{Fendy}.   A notable difference between  FENDy and the XF framework introduced by  Abedi, Maitra and Gross \cite{amg10}, is that  the former  is  based on the complex scalar potential rather than on the vector potential.  The  XF  formalism  is based on the product form of the electron-nuclear wavefunction $\Psi(\bm R,q,t)$, 
\be \Psi(\bm R,q,t) = \psi(q,t) \Phi(\bm R,q,t),\label{eq:XFWF} \ee 
 where $\psi(q,t)$ is the purely {\it nuclear} component    and   $\Phi(\bm R,q,t)$ is  the {\it electronic} component.    
The latter satisfies the  partial normalization condition,  i.e. it is normalized to one for any configuration of the nuclei, 
\be \bra \Phi|\Phi \ket_{\bm R} = 1 \mbox{~for~all~} q \mbox{~and~} t.  \label{eq:normalization}\ee
For Eq. (\ref{eq:XFWF}) to be formally exact, the 'electronic' component $\Phi$ is a function  of all spatial coordinates,  $\bm R$, $q$,   and  of time, $t$,  just as the full molecular wavefunction $\Psi$.    
Yet, the XF representation  of  $\Psi$ offers the following potential advantages.  If the nuclear  dynamics is largely captured by  $\psi(y,t)$,  then  (i) the resulting electron-nuclear terms might be small and, thus,  amenable to approximations. (ii) The nuclear dynamics may be treated approximately,   for example, using an ensemble of  semiclassical or quantum trajectories. Such description of the nuclear DOFs  may allow to  circumvent the formal exponential scaling with the system size even for large amplitude motion,  while retaining the dominant NQEs.  (iii)  The resulting time-dependent Schr\"odinger equation  (TDSE) on $\Phi$, coupled to propagation of $\psi$,  could be solved  without explicit reference to the electronic eigenstates,  treating  $\Phi$  as a wavepacket.  

The nuclear and electronic TDSEs are obtained by adding and subtracting  a complex  time-dependent potential energy surface (TDPES) $V_d(q,t)$  to the TDSE on  the molecular wavefunction which  separate as follows: 
\be \hat{K}_q\psi+V_d(q,t)\psi=\imath \frac{\pd\psi}{\pd t} \label{eq:nucTDSE},\ee
\be \hat{H}^{BO}\Phi+(\hat{D}^{(2)}+\hat{D}^{(1)})\Phi +\hat{V}^{ext}\Phi -V_d(q,t)\Phi=\imath \frac{\pd\Phi}{\pd t} \label{eq:eTDSE}.\ee 
The nuclear hamiltonian in Eq.  (\ref{eq:nucTDSE}),  composed  of  the nuclear kinetic energy, $\hat{K}_q$ and the so-far unspecified complex TDPES, $V_d$, 
\be V_d(q,t):=V_r(q,t)+ \imath V_i(q,t),  \label{eq:Vd} \ee
  does not include  any  explicit electron-nuclear  coupling terms. 
The electronic TDSE (\ref{eq:eTDSE})  includes   the Born-Oppenheimer BO Hamiltonian,     $\hat{H}^{BO}$,  of  Eq. (\ref{eq:HBO}),   consisting of  the  electronic kinetic energy and all Coulomb interactions, and  if present,  the external field $\hat{V}^{ext}$.  This TDSE also    includes the coupling terms,   $\hat{D}^{(1)}$ and $\hat{D}^{(2)}$,  containing the derivatives of $\Phi$ with respect to the nuclear coordinate,
\be \hat{D}^{(1)}:=-\frac{\grad_q\psi}{\psi}\frac{\grad_q}{M} \label{eq:D1},\ee~
\be \hat{D}^{(2)}:=-\frac{\grad^2_q}{2M}. \label{eq:D2}\ee  The derivative coupling operator,   $\hat{D}^{(1)}$,  involves  
 the log-derivative of $\psi$,  $\grad_q \psi/\psi$,    presenting   challenges for the numerical implementation of  the  XF equations.  
 For future reference, 
 let us define this log-derivative  through the nonclassical ($r_\psi$)  and classical ($p_\psi$)  components  of the {\it nuclear momentum operator} applied to the nuclear  function, 
 \bea r_\psi(q,t): & = & \Re(\psi^{-1}\grad_q\psi) =  \frac{\grad_q|\psi|}{|\psi|}, \label{eq:r_psi} \\
 p_\psi(q,t): &  = & \Im(\psi^{-1} \grad_q\psi) = \grad_q (\arg{\psi}).\label{eq:p_psi}
 \eea 
  and  define their counterparts   for  the electronic wavefunction, 
  \bea r_\Phi(\bm R,q,t): & = & \Re(\Phi^{-1}\grad_q\Phi) =  \frac{\grad_q|\Phi|}{|\Phi|}, \label{eq:r_Phi} \\
  p_\Phi(\bm R,q,t): &  = & \Im(\Phi^{-1} \grad_q\Phi) = \grad_q (\arg{\Phi}).\label{eq:p_Phi}
  \eea 
 
  Now let us turn to the TDPES, or the nuclear potential defined in  Eq. (\ref{eq:Vd}),   which governs  the dynamics of $\psi(q,t)$,   and thus the factorization of $\Psi(\bm R,q,t)$. 
  The choice of   $V_r$ should minimize  the average residual nuclear momentum in $\Phi$,   
  \be \overline{p_\Phi}:= \bra \Phi|p_\Phi|\Phi\ket_{\bm R},\label{eq:p_phi_av}\ee 
  while the  role of $V_i$ is to maintain the partial normalization condition, Eq. (\ref{eq:normalization}).  
 As shown in Ref. \cite{Fendy}  the norm-conserving  $V_i$  is equal to
\be V_i(q,t)  =  -\frac{1}{M}(2r_\psi+\grad_q)\overline{p_\Phi} \label{eq:exactVi}, \ee 
and there is a condition on the  'ideal' $V_r$, i.e. such $V_r$ that $\overline{p_\Phi} = 0$, 
\be  \grad_q V_r(q,t)  = \bra \Phi|\grad_q (\hat{H}_{BO}+\hat{V}^{ext})|\Phi \ket_{\bm R}+\left(4r_\psi+2\grad_q\right)
\bra \Phi|\hat{D}^{(2)}|\Phi \ket_{\bm R}. \label{eq:idealVr}\ee 
If   this  condition is fulfilled, then $\overline{p_\Phi}=0$  and $V_i\equiv 0$.     
 While in one nuclear dimension  $V_r$ can be reconstructed  (up to a time-dependent constant) from $\grad_q V_r$  of   Eq. (\ref{eq:idealVr}), this is not the case for  the coupled  nuclear motion in  many dimension. The ambiguity in the choice of $V_r$ is related to the ambiguity of separating the full molecular phase between the nuclear and electronic components.  Some choices of  $V_r$ are discussed in Ref. \cite{Fendy}.  
  
  The  imaginary potential $V_i$ of Eq. (\ref{eq:exactVi}) is exact if $\Phi$ is described in a complete basis. To maintain normalization of $\Phi$ in a finite basis used in this work,  we  
   introduce the following  definition of   $V_i$,  
  \be V_i = \Im(\bra \Phi|\hat{H}_{el}| \Phi\ket_{\bm R}),   \label{eq:newVi}\ee
   compatible with the hermiticity of the full electronic hamiltonian,
  \be \hat{H}_{el}= \hat{H}_{BO}+\hat{D}^{(1)}+\hat{D}^{(2)}+\hat{V}^{ext},  \label{eq:He} \ee 
   once it is expressed in the matrix form.  Both definitions of   $V_i$ conserve  the trajectory weights and become equivalent in the complete basis limit. 
   
   The real potential $V_r$, which is the  effective nuclear potential driving the time-evolution of the  nuclear wavefunction $\psi$, also needs to be modified from Eq. \ref{eq:idealVr} to account for finite electronic basis.  We use a convenient definition  similar to Eq. (\ref{eq:newVi}):
   \be V_r= \Re(\bra \Phi|\hat{H}_{el}| \Phi\ket_{\bm R}) \label{eq:newVr}.  \ee 
   While not strictly minimizing $\overline{p_\Phi}$, this is essentially a mean-field potential associated  with $\hat{H}_{el}$, the latter comprised of the Born-Oppenheimer hamiltonian, external  field, and electron-nuclear coupling.  Examples of  other choices  of $V_r$  can be found in Ref. \cite{Fendy};  construction  of optimal $V_r$ for general systems is deferred to future work.  

\subsection{Quantum trajectory (QT)  formalism} \label{sec:QTs}
The de Broglie-Bohm-Madelung, often referred to as  'Bohmian',  formulation of quantum mechanics \cite{bohm52,wyatt_book}  provides an  alternative to  conventional basis/grid representations of  multidimensional nuclear wavefunctions by recasting the  TDSE  into the trajectory framework.  This is conceptually appealing  because  of  an intuitive  connection to  classical molecular mechanics, widely used to study high-dimensional molecular  systems.  A review of the QT  field is beyond  the  scope of this paper, but let  us   mention that  the  QTs   were used  to simulate  single-surface  and nonadiabatic  dynamics  
    \cite{Bohmian,meier2023,Dupuy22a,Lombardini24,wyatt01a,Dupuy22b,garashchuk06b}, as well as  to  interpret the quantum phenomena  \cite{sanz07,angel2024}. The Bohmian framework  also   served   as a starting point  for approximate and semiclassical  methods, e.g.  linearized quantum force \cite{garashchuk04}, quantized Hamiltonian dynamics \cite{prezhdo06,akimov2018}, the QT-guided adaptable gaussian bases \cite{QTAG}, and  for  the QT surface-hopping methods  with and without the XF of molecular wavefunctions  \cite{QTSH,AkimovQTSH,XFTullyModels}.

Within the Bohmian formulation 
a wavefunction is represented as an ensemble of quantum trajectories (QTs), evolving in time according to the classical-like equations of motion. The time-evolution of the  probability density obeys the continuity equation.  All the 'quantumness' comes from  the  non-local quantum potential  added to the external 'classical' potential. 
In this section the formalism is outlined using the notations consistent with the rest of the paper, namely: one-dimensional coordinate $q$, the particle mass $M$ and an external,  possibly time-dependent,  potential $V(q)$.   A generalization to many dimensions and non-Cartesian coordinates  is given, for example, in Ref. \cite{garashchuk05a}.
  
The Bohmian  equations of motion follow from  the polar representation of a complex wavfunction,  $\psi(q,t)$,   in terms of its real phase, $s(q,t)$, and amplitude, $|\psi(q,t)|$,  
 \be \psi(q,t):=|\psi(q,t)|e^{\imath s(q,t)}, \label{eq:QTWF} \ee  
 substituted into the TDSE. 
 We will use  subscript $t$ to indicate {\it quantities associated with the  QT} at a position $q_t$,  as opposed to  {\it functions of $q$}. 
 The gradient of the wavefunction phase,  $p(q,t)=\grad_q s(q,t)$,  
 evaluated at 
 the QT position $q_t$ defines the QT momentum,  
 \be p_t= p(q,t)|_{q=q_t}, \label{eq:pt}\ee
   which determines the dynamics of  $q_t$  according to the classical equation of motion (EOM), 
 \be \frac{dq_t}{dt}= \frac{p_t}{M}. \label{eq:qdot}\ee
 The  trajectory velocity, $p_t/M$,  defines the Lagrangian frame of reference, 
 \be \frac{d}{dt}:=\frac{\pd}{\pd t}+\frac{p_t}{M}\grad_q.   \label{eq:LagFrame} \ee
  In this QT-frame of reference  the EOMs following from the TDSE are:  
\bea \frac{dp_t}{dt} &=& -\grad_q(V+U)|_{q=q_t} \label{eq:pdot}
\\ \frac{ds_t}{dt}&=&\frac{p_t^2}{2M}-(V+U)|_{q=q_t}, \label{eq:dsdt}
\\ \frac{d\rho_t}{dt}&=&-\frac{\rho_t}{M}\left.(\grad_q p)\right|_{q=q_t},\label{eq:drhodt}
\eea
 where $\rho_t=|\psi(q_t)|^2$ is the probability density  and   $U(q,t)$ is the quantum potential, 
 \be U(q,t):=-\frac{\hbar^2}{2M}\frac{\grad^2_q|\psi|}{|\psi|}. \label{eq:qpot} \ee
The latter is $\hbar^2/M$ ($\hbar$ is included here explicitly)   function which   includes exactly  the  NQEs into otherwise classical Eqs (\ref{eq:qdot}-\ref{eq:dsdt}).     

A notable property of the QT dynamics is conservation of  the trajectory weight,  $w_t$, 
\be w_t:=w(q_t)= |\psi(q_t)|^2dq_t, \label{eq:weight} \ee 
defined as the probability density within the volume element $dq_t$ associated with a trajectory,  $q_t$. 
As follows from the QT equations and Eq. (\ref{eq:drhodt}) \cite{gr20},  $w_t$ 
is constant in time, $$w_t=w_0=w,$$   allowing one to  compute the  expectation values of position-dependent and some momentum-dependent operators, 
\be \bra \hat{O} \ket_t = \int O(q) |\psi(q,t)|^2 dq \approx \sum_{i=1}^{N_{traj}} O(q_t^{(i)})w^{(i)} \label{eq:SumoverTraj},\ee 
without explicit construction of $\psi(q,t)$. In other words,   
upon the initial discretization of $\psi(q,0)$ in terms of QTs and the associated weights,  $\bra \hat{O}\ket_t$   can be  evaluated as a simple weighted sum over the  trajectories at any time $t$.  
  Equation (\ref{eq:SumoverTraj})  is used in this work when appropriate.   

\subsection{Equations of motion}
Here we define the EOMs for the QTs evolving according to  the FENDy formalism and for the electronic expansion coefficients,  given  the  expansion of $\Phi(\bm R, q, t)$ in a stationary  electronic basis, $\{\phi(\bm R, q)\}$, 
\be \Phi({\bf R},q,t):=\sum_{\mu=1}^{N_{bas}}C_\mu(q,t)\phi_\mu({\bm R},q) \label{eq:eansatz}. \ee
Index   $\mu$ enumerates  the  functions in the   basis   $\{\phi\}$ of the  size $N_{bas}$. 

In FENDy, the nuclear trajectory momentum is defined by the {\em total} nuclear momentum: 
\be p_t:=p_\psi + \overline{p_\Phi} \label{eq:TrajMom}, \ee
which includes the residual nuclear momentum of the electronic wavefunction, $\overline{p_\Phi}$ (Eq. (\ref{eq:p_phi_av})). Thus, the EOM for the nuclear wavefunction momentum is modified accordingly  by  the addition of a Lagrangian term proportional to $\overline{p_\Phi}/M$: 
\be \frac{dp_\psi}{dt}= -\left.\grad_q (V_r+U)\right|_{q=q_t}+\frac{\overline{p_{\phi}}}{M}\grad_q p_\psi \label{eq:dppsidt}. \ee
 In the numerical implementation we use an equivalent to Eq. (\ref{eq:qpot}) representation of   $U$  in terms of $r_\psi$,  
 \be U(q,t):=-\frac{r_\psi^2+\grad_q r_\psi}{2M}.\label{eq:Ur}\ee
 The trajectory positions evolve according to the total nuclear velocity $p_t$/M (Eq. (\ref{eq:TrajMom})) as  $dq_t/dt = p_t/M$.
The trajectory weights are conserved due to the imaginary potential, $V_i$, as proved in Ref.  \cite{Fendy}. The non-classical momentum itself is  evaluated on the fly from the distribution of the trajectories and their weights as described in the next section. 

The time evolution for the electronic basis coefficients,  obtained from Eq. (\ref{eq:eansatz}) and Eq. (\ref{eq:eTDSE}) in the moving frame,  is governed by  the following EOM,  
\be \frac{d\bm{C}(q,t)}{dt}=-\imath \mathbf{S}^{-1}\mathbf{H}_D\bm{C} +\frac{p_t}{M}\grad_q\bm{C} \label{eq:dcdt}. \ee
In Eq. (\ref{eq:dcdt})  $\mathbf{H}_D$ denotes the full electronic hamiltonian $\hat{H}_{el}$ in the matrix form with subtraction of $V_d$,  
\be \mathbf{H}_D = \mathbf{H}^{BO}+\mathbf{V}^{ext}
-\frac{\grad_q\psi}{\psi}\frac{1}{M}\left(\mathbf{D}^{(1)}+{\mathbf S}\grad_q\right)
-\frac{1}{2M} \left(\mathbf{D}^{(2)}+{\mathbf S}\grad^2_q+2\mathbf{D}^{(1)}\grad_q\right)
-{\mathbf S}V_d \label{eq:BigElecHam},  \ee
 with the following matrix elements, where $(\mu,\lambda)=[1,N_{bas}]$:  
\bea
{S}_{\mu\nu} & = & \bra \phi_\mu | \phi_\lambda \ket_{\bm R}, \label{eq:Smat} \\
{H}^{BO}_{\mu\nu} & = & \bra \phi_\mu |\hat{H}^{BO} \phi_\lambda \ket_{\bm R}, \label{eq:HBOmat} \\
{V}^{ext}_{\mu\nu} & = & \bra \phi_\mu |\hat{V}^{ext} \phi_\lambda \ket_{\bm R}, \label{eq:Vextmat} \\
D^{(1)}_{\mu\nu} & = & \bra \phi_\mu| \grad_q \phi_\lambda \ket_{\bm R}, \label{eq:D1mat} \\
D^{(2)}_{\mu\nu}\ & = & \bra \phi_\mu| \grad^2_q \phi_\lambda \ket_{\bm R}. \label{eq:D2mat} 
\eea
In Section \ref{sec:Implementation} the vectors  $\bm{C}(q,t)$,  defined  for each trajectory  $q_t$, will be listed together  as columns of  the matrix ${\mathbf{C}}$, its size  being $N_{bas}\times N_{traj}$. 

\subsection{ Basis set projection within the QT framework} \label{sec:BasisProjections}
Following  Ref. \cite{garashchuk04} the necessary for the dynamics  derivatives of functions,  known  at the trajectory positions, such as expansion coefficients ${\mathbf{C}}$ and  nuclear momenta $p_\psi$,  are computed from their projections on a basis  $\bm f=(f_1(q),f_2(q),\dots,f_{N}(q)$)  of the size $N$.  Let  $F(q)$ and $\tilde{F}(q)$   denote a function/property and its expansion in a basis, respectively.   Using $\bm b=(b_1,b_2,\dots,b_{N})$  as a vector of the expansion coefficitens,  $\tilde{F}(q)$  is given by     
\be \tilde{F}(q)=\sum_{k=1}^{N} b_{k}f_{k}(q) =\bm f\cdot \bm b. \label{eq:BasisProj} \ee
The derivatives  of $F(q)$  are then approximated by analytic derivatives  of Eq. (\ref{eq:BasisProj}). 
As the basis functions are analytic, this approach is useful for differentiation on an unstructured grid and as a low-pass filter suppressing noise associated with the grid sparsity.  

For the quantities required to solve the nuclear TDSE (\ref{eq:nucTDSE}),  the expansion coefficients $\bm b$   are determined from the Least Squares Fit (LSF)  weighted by the nuclear density \cite{garashchuk04}.   Taking advantage of the QT weight conservation (Eq. (\ref{eq:SumoverTraj})),    the LSF is determined by minimizing $I$ evaluated  as a weighted sum over trajectories, 
\be I=\bra \psi(q)|(F(q)-\tilde{F}(q))^2|\psi(q)\ket \approx \sum_{i}^{N_{traj}} \left( F(q_t^{(i)})-\tilde{F}(q_t^{(i)}) \right)^2 w^{(i)} \label{eq:LSF}. \ee 
 Minimization of $I$ with respect to the elements of $\bm b$    gives their optimal values  as solutions to   the  linear matrix  equation,
\be {\mathbf \Omega}\bm{b}=\bm{d}. \label{eq:LSFEq} \ee
Here ${\mathbf \Omega}$ is the overlap matrix of the basis functions and 
$\bm d$ is the vector of projections with the elements, 
\bea
\Omega_{kk'} & = &  \sum_{i}^{N_{traj}} f_k(q_t^{(i)}) f_{k'}(q_t^{(i)})w^{(i)},\label{eq:over_kk}\\
d_{k}& = & \sum_{i}^{N_{traj}} F(q_t^{(i)})f_{k}(q_t^{(i)})w^{(i)}. \label{eq:rhs_k}\eea
Equation (\ref{eq:LSFEq}) is solved directly for  $\bm b$, which defines $\tilde{F}(q)$ used to compute the necessary derivatives. 

For the quantities in the electronic TDSE (\ref{eq:eTDSE}) --   the coupling terms  $\hat{D}^{(1)}$ and $\hat{D}^{(2)}$ and the residual nuclear momentum  $\overline{p_\Phi}$ --  we need the  derivatives of the XF expansion coefficients arranged as the matrix ${\mathbf{C}}$.  These are evaluated    by the same  procedure as above (Eqs (\ref{eq:LSF}-\ref{eq:rhs_k})) with  exception of  using {\it equal weights}  in Eq. (\ref{eq:LSF}), i.e.  $w_i=1$ for $i=[1,N_{traj}]$.  This  gives a statistical LSF, rather than the one based on integration over the nuclear space.
Excluding the nuclear density gives more accurate fitting of the elements of  ${\mathbf{C}}$ in the regions of low wavefunction probability, reducing the error propagation from the low to high molecular density  regions in the course of dynamics. 

\section{Computational algorithms and implementation} \label{sec:Implementation}
The FENDy approach described in Section \ref{sec:theory} and the electronic structure basis sets are implemented in Maple \cite{maple}.  In this section we provide an overview of the implementation summarized in Tables  \ref{tab:Algorithim} and \ref{tab:TimeLoop}. 
The three major blocks include 
\begin{enumerate}[label=({\bf \roman*}):]
    \item initialization of the simulation parameters, projection and electronic bases (Table \ref{tab:Algorithim}, steps 1-5);
    \item initialization of the wavefunction, trajectories  and electronic matrix elements (Table \ref{tab:Algorithim}, steps 6-8);
    \item time-integration (Table \ref{tab:TimeLoop}, steps (9-23)).
\end{enumerate}

\begin{table}[]
    \centering
    \begin{tabular}{ll}
    \hline
   \multicolumn{2}{c}{\bf FENDy algorithm} \\ 
    \hline
   \multicolumn{2}{l}{ */\textbf{Initialize basis sets}*/} \\
    1:~~ & Initialize standard electronic structure basis set \\
   \multicolumn{2}{l}{  */\textbf{Read Parameters}*/} \\
    2:~~ & QT parameters: \\
   & Number of quantum trajectories and span \\
    3:~~ &  Electronic projection basis parameters: \\
   & Number of Fourier functions and carrier frequency  \\
    4: & Laser parameters: \\
  &  Width, peak time, intensity and frequency of the laser (Table \ref{tab:ModelParams}) \\
  \multicolumn{2}{l}{ */\textbf{Initialize auxiliary bases}*/} \\
    5:&  Define arrays for projections of $p_\psi$,$r_\psi$,$V_d$, and $\mathbf{C}$ \\
   \multicolumn{2}{l}{*/\textbf{Initialize trajectories and electronic wavepackets}*/} \\
    6: & Compute electronic matrix elements in AO basis \\
    7: & Initialize trajectories on a uniform grid \\
    8: & Initialize interpolation scheme for electronic matrix elements \\
\multicolumn{2}{l}{ */\textbf{Time loop}*/} \\\hline
    \end{tabular}
    \caption{Steps of  the  numerical implementation of  FENDy with complex potential. The time loop is described in Table \ref{tab:TimeLoop}. The auxiliary basis sets with the exception of the Fourier basis are pre-defined. The electronic matrices (Eqs (\ref{eq:Smat}-\ref{eq:D2mat})) are pre-computed on a grid in $q$ and are interpolated during the propagation.}
    \label{tab:Algorithim}
\end{table}

The projection basis sets for the nuclei quantities are preset as follows:
\begin{itemize}
    \item[--]   the nuclear momentum  basis is  $\bm f^{p_\psi}=\{1,e^{-3 q},\text{erf}(q/10)\}$;
    \item[--]   the non-classical momentum basis  is  $\bm f^{r_\psi}=\{1,q,e^{-3 q}\}$; 
    \item[--]  the basis for $V_d$ (Eq. (\ref{eq:Vd})) is $\bm f^{V_d}=\{1,e^{-11q/20},e^{-11q/10 },1/q\}$.  
\end{itemize}
In block ({\bf i}) the  basis $\bm f^{r_\psi}$  is chosen according to  the analysis  
   of the QT  evolution of a  wavepacket  in the Morse  potential
\cite{garashchuk06a}.   The basis $\bm f^{p_\psi}$ is chosen  to describe a constant momentum for $q\rightarrow\infty$ with the parameters approximating the split-operator result, while  the basis  $\bm f^{V_d}$ is determined by the fitting of the  ground and excited Born-Oppenheimer  surfaces. The electronic expansion coefficients ${\mathbf{C}}$ (Eq. (\ref{eq:eansatz}))  are projected on the Fourier basis   of  $N_p$  sine and $N_p$ cosine functions  with a carrier frequency of $\pi/2$,  plus a constant, resulting in a total basis  of $N=2N_p+1$ functions.  The electronic basis $\{\phi\}$ is defined  either as the  standard electronic structure basis set such as  6-31G,    henceforth referred to as  {\it "atomic orbital basis"} (AO), or as the eigenfunctions of $\mathbf{H}^{BO}$ henceforth referred to as  {\it "eigenbasis"}.  In block ({\bf ii}) for computational  efficiency  the electronic matrix elements (Eqs (\ref{eq:Smat}-\ref{eq:D2mat})) are pre-computed on a grid in $q$. For the dynamics in the  eigenbasis representation,  ${\mathbf H^{BO}}$ is  diagonalized and the coupling matrices are computed in the eigenbasis. During the dynamics the pre-computed matrices are interpolated for any $q$ using four-point Lagrange scheme. Details of the interpolation are described in Appendix \ref{sec:Interp}. 
\begin{table}[]
    \centering
    \begin{tabular}{ll}
    \hline
    \multicolumn{2}{c}{\bf Time propagation algorithm} \\
    \hline
    \multicolumn{2}{l}{*/\textbf{Time loop}*/} \\
  \multicolumn{2}{l}{  for $n$ from 1 to  $N_{steps}$ {\em do}} \\
    \multicolumn{2}{l}{~~~*/\textbf{Self-consistent propagation}*/}\\
    ~~{\em do} & \\
    \multicolumn{2}{l}{~~~*/\textbf{Compute time-derivatives}*/}\\
    ~~~9:~~  & Electronic matrix elements via interpolation \\
    ~~~10:~~ & Project ${\mathbf{C}}$ into Fourier basis \\
    ~~~11:~~ & Compute $(\hat{D}^{(1)}+\hat{D}^{(2)})|\Phi\ket$ \\
    ~~~12:~~ & Compute $\overline{p_\Phi}$ \\
    ~~~13:~~ & Compute $\frac{d}{dt}\mathbf{C}$,$\bra \Phi | \mathbf{H}_D | \Phi \ket_{\bm{R}}$ \\
    ~~~14:~~ & Project $V_d$ into $\bm{f}^{V_d}$\\
    ~~~15:~~ & Project $p_{\psi}$ into $\bm{f}^{p_\psi}$\\
    ~~~16:~~ & Compute quantum force \\
    ~~~17:~~ & Compute $\frac{d}{dt}{p}_\psi$ and $\frac{d}{dt}{q}_t$ \\
    ~~~18:~~ & Compute sum of squares change in $F(t+dt)$ from previous step \\
    ~~~19:~~ & Compute average of $F(t+dt)$ and $F(t)$ \\
    ~~~20:~~ & Project $r_\psi$ of average into $f^{r_{\psi}}$ \\
    \multicolumn{2}{l}{~~~{\em until} \texttt {error < threshold} ~{\em  or}~ \texttt{iteration > max(iteration)} }\\
    \multicolumn{2}{l}{~~~*/\textbf{Low pass filter}*/}\\
    ~~~21:~~ & Project ${\mathbf{C}}$, replace with $\tilde{\mathbf C}$  and re-normalize. \\
    ~~~22:~~ & Project $r_\psi$ and  replace with $\tilde{r}_\psi$ \\
    ~~~23:~~ & Project $p_\psi$  and  replace with $\tilde{p}_\psi$ \\
    \multicolumn{2}{l}{{\em end do}}
      \end{tabular}
    \caption{The time loop with self-consistent propagation scheme (steps 9-20).  The low pass filter (steps 21-23) is applied at the end of each time-step.}
    \label{tab:TimeLoop}
\end{table}

The time integration of block ({\bf iii}) is performed via a self consistent loop  comprised of steps 9-20 in Table \ref{tab:TimeLoop},   modeled  after  Ref. \cite{jakowski2025}. Essentially, the time derivatives of all functions $F(t)$ --  the trajectory positions $\{q_t\}$, momenta $\{p_t\}$, and electronic expansion coefficients ${\mathbf{C}}$ -- are evaluated using the average of their initial value $F(t)$ and value at the next step $F(t+dt)$ until consistency in $F(t+dt)$ is achieved.  
The  error of the self-consistent loop is  equal to
$$\texttt{error} = \sum_{i=1}^{N_{traj}}\left( (q_t^{(i)})^2+(p_t^{(i)})^2+\sum_{\mu=1}^{N_{bas}} |C_{i\mu}|^2\right).$$
 The projection of ${\mathbf{C}}$ is outlined in Section \ref{sec:BasisProjections}.  
 The quantities required to update ${\mathbf{C}}$ are computed in steps 11-13, using 
 $ \psi^{-1}\grad_q \psi=r_\psi+\imath p_\psi$ in evaluation of $\hat{D}^{(1)}$. 
 Step 12 is the computation of the residual nuclear momentum, which  
  is equal to  
  $$\overline{p_\Phi}:=-\imath \bra \Phi \grad_q \Phi \ket,$$ provided that  $\Phi$  satisfies the partial normalization condition Eq. (\ref{eq:normalization}).  
Therefore, $\overline{p_\Phi}$ is computed directly from $\bra\Phi|\hat{D}^{(1)}\Phi \ket_{\bm{R}}$. 
Since  the low-pass filter replaces $\mathbf C$ with its projected approximation, $\tilde{\mathbf C}$,   to maintain the normalization condition, the enusing  $\Phi$ is re-normalized. The re-normalization introduces certain errors  and, ideally, a constrained fitting procedure which  enforces normalization while keeping ${\mathbf{C}}$ in a basis, is desirable. Such a procedure (not discussed here) is  relatively straightforward for an orthonormal electronic basis. However, we found this approach to be computationally expensive and not worth the effort in this application. 

The projections of $V_d$ and $p_\psi$ (steps 14, 15 and 22) follow the procedure in Section \ref{sec:BasisProjections} using their respective bases. In step 16 the quantum force is computed from   $\grad_q U$  of Eq. (\ref{eq:Ur}), necessary to update the QT  momenta (along with the QT  positions) in step 17. 
In the LSF procedure  (Eq. (\ref{eq:LSFEq}))  for  $r_\psi$,  the division 
by $|\psi(q,t)|$ is avoided by  using integration by parts \cite{garashchuk03a} when  computing  $\bm{d}$,  
\be d_{k}=-\frac{1}{2}\bra \psi| \grad_qf_{k} |\psi\ket_q=-\frac{1}{2}\sum_{i}^{N_{traj}}  w^{(i)} \left.\grad_q f_{k}\right|_{q=q_t^{(i)}}. \ee
Once  the self-consistency error is below the  threshold  (or maximum  number of iterations is  reached),   $\mathbf{C}$, $p_\psi$ and $r_\psi$ are projected on their respective bases  once more in steps 21-23. This "low pass filter" is performed, i.e. these quantities are replaced with their fitted counterparts,   to mitigate the high-frequency 'noise' acquired during the approximate propagation.  The accuracy of the dynamics is achieved by increasing the Fourier basis (and improving the  projection bases for $p_\psi$ and $r_\psi$) at the cost of smaller time-step and/or more iterations in the self-consistent cycle.   

\section{Results and Discussion}\label{sec:results}
The results discussed below are obtained for the 6-31G electronic basis set.  Its  functions  act as the basis functions  $\phi_\mu$ of Eq. (\ref{eq:eansatz}) when the formalism is implemented in  the AO basis. For the eigenbasis implementation,   $\phi_\mu$  are defined as the eigenfunctions of  $\mathbf{H}^{BO}$  evaluated  in the 6-31G AO basis.   Unless noted otherwise.
we discuss the results computed in the eigenbasis. 
  The performance of  FENDy   is compared  to the  dynamics of  the nuclear wavefunction  propagated in time  employing  the conventional  split-operator/Fast-Fourier Transform  on a grid \cite{FFS82,kosloff88}.   
 In the eigenbasis  implementation  the couplings beyond the ground and first excited electronic states were found to be very small and, thus, they are set to zero to  speed up the calculations and to facilitate comparison with the split-operator dynamics obtained for the same two-state nonadiabatic  Hamiltonian in the presence of the laser field,  $V^{ext}$.  The split-operator results are referred to 'exact'  throughout this section.   The  initial nuclear wavefunction is taken as a gaussian,
 \be \psi(q,0)=\left(\frac{\alpha_{coh}}{\pi}\right)^{1/4}\exp\left(-\frac{\alpha_{coh}}{2}(q-\bra q\ket_{0})^2\right), \label{eq:psi0}\ee  matching the ground state of the harmonic approximation to the ground BO surface at the minimum computed in the STO-3G basis.  The ground electronic states is the only state populated at time $t=0$.   The  results are presented for the  dynamics parameters  listed in Table \ref{tab:num_params} unless noted otherwise.    

\begin{table}[]
    \centering
    \begin{tabular}{|c|c|c|c|c|}
        \hline
        $N_{traj}=50$ & $\rho_{min}=10^{-8}$ &  $dt=0.25$ a.u. &  \texttt{threshhold}=$10^{-15}$ & \texttt{iterations}=$10$ \\\hline $d_{\omega0}=\pi/2$ & $N_p=12$ &
        $\epsilon=10^{-9}$ & {$\alpha_{coh}$=8.16} & {$\bra q\ket_0$=2.004}  \\\hline
         \end{tabular}
    \caption{The simulation parameters include number of trajectories ($N_{traj}$),  nuclear density cutoff ($\rho_{min}$) for $q_0^{(i)}$, time-step  ($dt$),self-consistent propagation convergence criterion \texttt{threshold} and \texttt{iterations} ,  carrier frequency ($d_{\omega0}$) and the number of sine/cosine functions ($N_p$) for the Fourier basis, unless specified otherwise, and interpolation error tolerance ($\epsilon$). The nuclear wavefunction  $\psi(q,0)$  is specified  by the width  $\alpha_{coh}$ and center $\bra q\ket_0$. 
    }
    \label{tab:num_params}
\end{table}

\begin{figure}
    \begin{tabular}{ll}
    a)  & b)   \\\vspace*{-.15in}
       \includegraphics[width=0.47\linewidth]{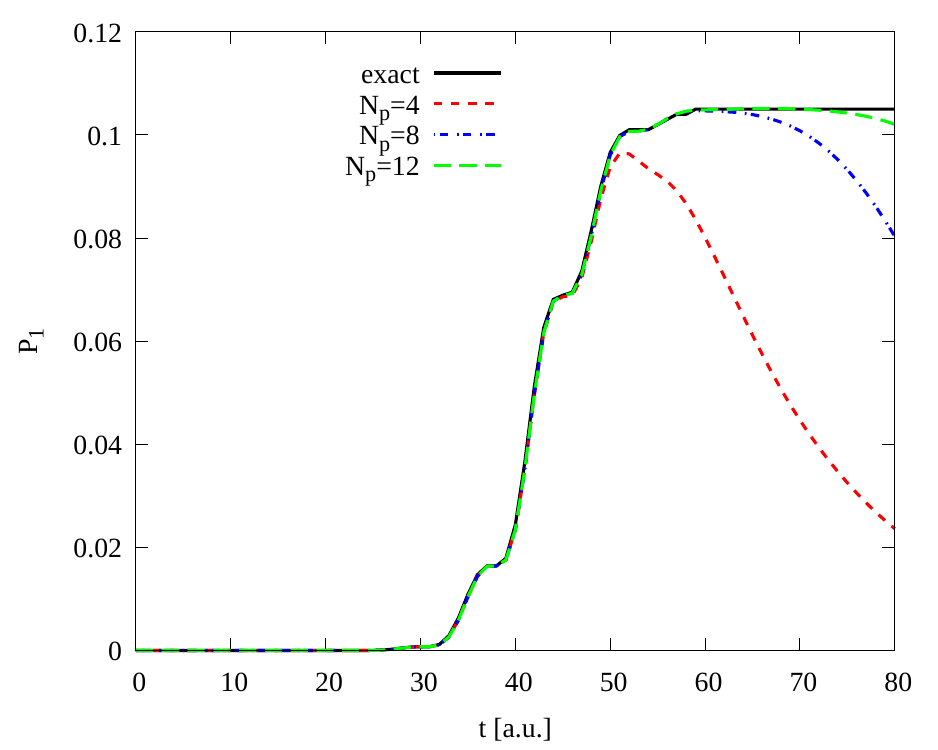} ~~~~    &  \includegraphics[width=0.47\linewidth]{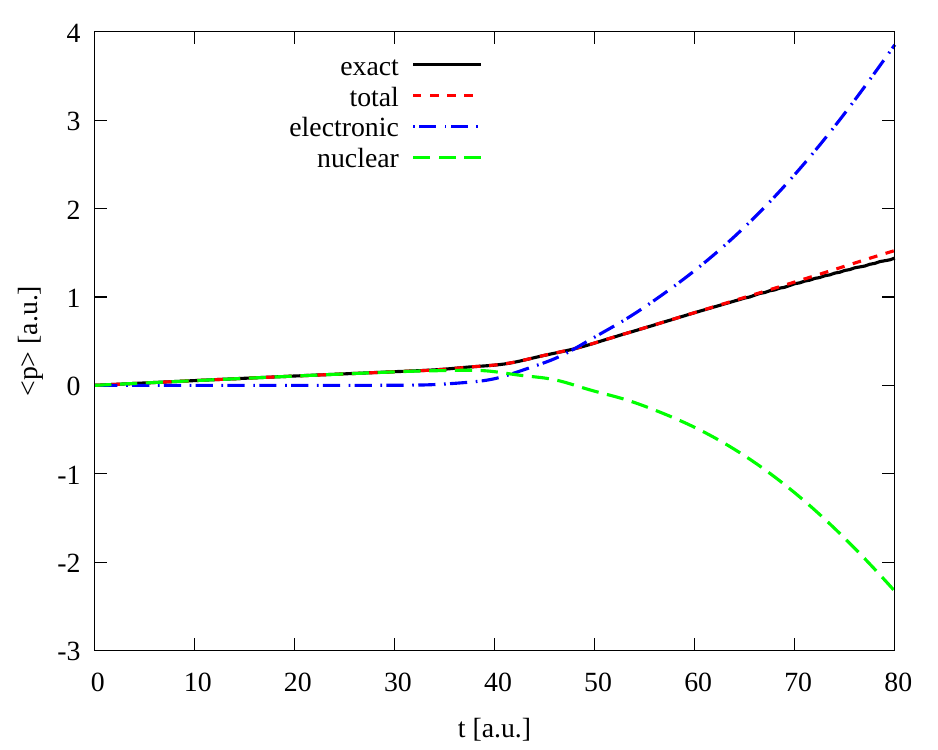}
        \end{tabular}
    \caption{(a) The  excited state populations, $P_1$,   obtained  with FENDy for $N_p=[4,8,12]$,  shown as red dash, blue dot-dash and green long dash, respectively, compared to the exact result (black solid line) as functions of time $t$.  (b) The total nuclear momentum  for the molecular wavefunction $\Psi$ ($\bra p\ket=\bra\Psi| \hat{p}|\Psi\ket$)  and its electronic ($\bra\psi|\overline{p_\phi}|\psi\ket_q$) and nuclear
    ($\bra\psi| p_\psi|\psi\ket_q$) components, shown as red dash, blue dot-dash  and green long dash, respectively,  as functions of time $t$.  The total momentum  $\bra p\ket$ from the exact dynamics is represented as black solid line.  } 
    \label{fig:UWPops}
\end{figure}

First of all let us examine how FENDy captures the population dynamics.  
Figure  \ref{fig:UWPops}(a) shows the  population of the excited electronic state, which was initially unoccupied,   for several values of the Fourier basis,  $N_p=\{4,8,12\}$, used for the electronic coefficients 
$\mathbf C$.   The population $P_1$ is defined as  average of the nuclear density weighted by the electronic excited state populations.
The population builds up   with the increase of the laser intensity, which peaks at $t_0=40$ a.u.,   in 'waves'  corresponding to the laser frequency.   The FENDy results  approach the exact results (black  solid line)  with the increase of the Fourier basis size. 
The  population 'loss'  at $t>50$ a.u., i.e. the drop-off of the population instead of reaching a plateau once the laser pulse is turned off,   is attributed to the re-projection of $\mathbf C$ into the basis at the 'low pass filter' block (steps 21-23 in Table \ref{tab:TimeLoop}). This trend is mitigated  by increasing  $N_p$.   
Once the laser intensity starts decreasing, the molecular wavefunction  exhibits dissociative behavior as can be inferred from the average nuclear momentum, $\bra p\ket$, displayed in Fig. \ref{fig:UWPops}(b). This quantity   grows slowly from its initial value of zero  up to $t_0$ and then exhibits notable increase,  which reflects the dissociative motion on the purely repulsive excited PES, once its population becomes appreciable reaching about $10\%$.   

\begin{figure}
    \begin{tabular}{ll}
    a)  & b)   \\\vspace*{-.15in}
      \includegraphics[width=0.45\textwidth]{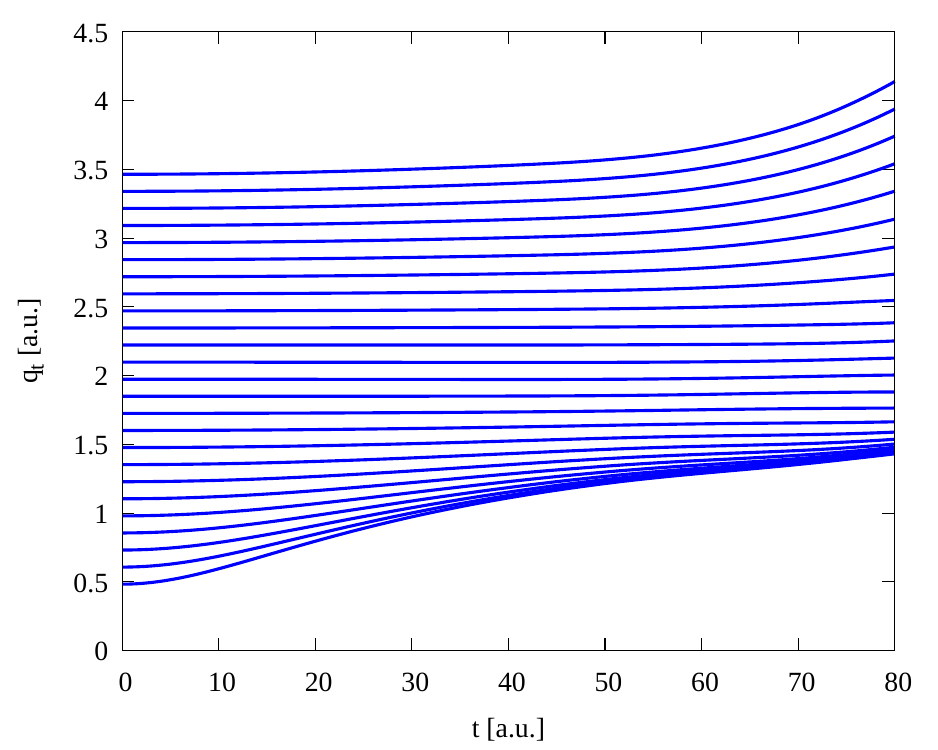}& \includegraphics[width=0.49\textwidth]{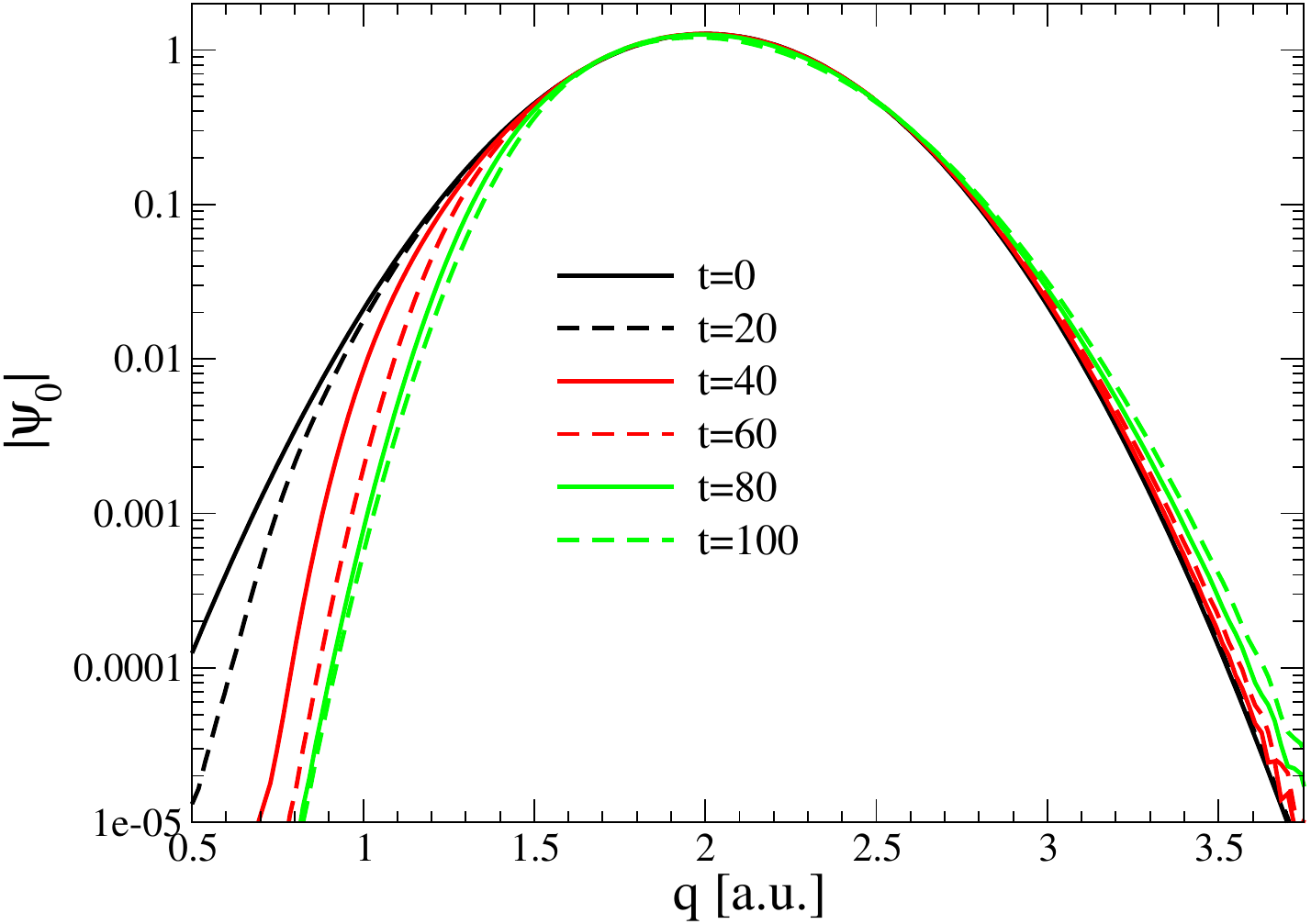}
        \end{tabular}
    \caption{Time-evolution of the nuclear wavefunction. (a) Positions $\{q_t\}$ of the quantum trajectory ensemble  representing 
    the nuclear wavefunction $\psi(q,t)$ within the XF representation as functions of time. Every other trajectory is shown for clarity.   (b) 
    The  'exact' ground state amplitudes, $|\psi_0(q,t)|$, as functions of the nuclear coordinate $q$, obtained   from  the  conventional two-state nonadiabatic dynamics at times indicated in the legend. } 
    \label{fig:traj}
\end{figure}

The overall dynamics of the  nuclear wavefunction $\psi(q,t)$,  however,  is dominated by the  population of the ground electronic state. Within the QT representation of $\psi(q,t)$ this is seen in the evolution of the trajectory ensemble displayed in Fig. \ref{fig:traj}(a).  
The snapshots of the amplitudes of the ground state nuclear wavefunction, $|\psi_0(q,t)|$,  from  the  conventional two-state nonadiabatic dynamics ('exact' split-operator calculations) are shown in Fig. \ref{fig:traj}(b) as functions of $q$ on the   vertical log-scale. 
The initially gaussian wavepacket is pulling away from the repulsive wall (small $q$), which is seen as compression of the QTs  and squeezing of $|\psi_0(q,t)|$  with time for $q<1.5$ a.u. 
This compression of QTs is the cause of the numerical 'noise', managed in our simulation by the low-pass filter, 
 due to unphysical crossing of the trajectories resulting in the 'noisy' behavior of wavefunction properties computed at the QT positions,  such as  $\mathbf C$.   Focusing now on  the right side of  $|\psi_0(q,t)|$ we observe only  small changes, i.e. the probability density in the ground state is largely stationary, despite the loss of the probability to the excited state.   Within the XF, however,  the nuclear wavefunction $\psi(q,t)$ represents the 'collective' dynamics on both ground and excited BO surfaces, which is seen in the dissociative  behavior of QTs for $q>2.5$ a.u.  as time progresses.

\begin{figure}
    \centering
    \includegraphics[width=0.5\linewidth]{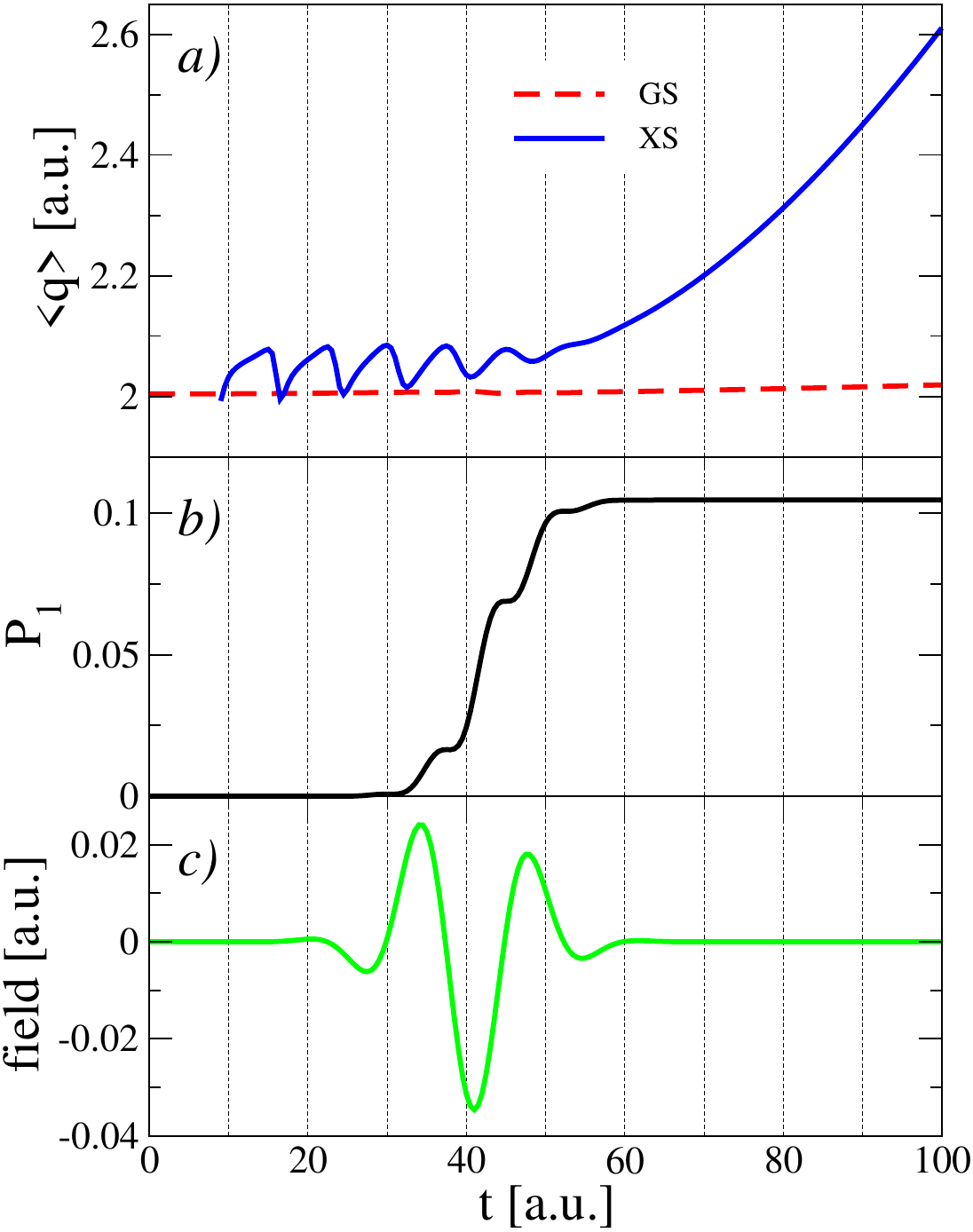}
    \caption{Properties of the excited state dynamics (computed exactly): (a) average position of the ground (red dashed line) and excited (blue solid line) nuclear wavefunctions;  (b) excited state population;  (c) the electric field of the laser pulse. }
    \label{fig:qav}
\end{figure}

Closer examination of the dynamics on the two BO surfaces, illustrated in Fig. \ref{fig:qav}, reveals that the dissociative dynamics begins after the laser intensity peaks, i.e.  for $t>t_0$.  The population on the excited state (blue line)  is effectively 'held in place' by the laser pulse during the build up of its population. This is  seen as  the oscillatory behavior of the average nuclear position, $\bra q\ket$,  on the excited state for $t<45$ a.u. and then increases as the  nuclear wavepacket starts to dissociate (blue line in Fig. \ref{fig:qav}(a)).   The average nuclear position on the ground state  (red dash) remains essentially unchanged for the time interval shown in the figure. The population of the excited state and the laser pulse as functions of time are  shown in Figs \ref{fig:qav}(b,c), respectively, to highlight the correlation between the population transfer  and the laser pulse.     
The  nuclear  wavefunction bifurcation associated with the dissociative dynamics  presents a major conceptual challenge to the XF formulation in general.  Within our implementation  it  is manifested in  the  singularities in $\mathbf{C}$,  which make  the numerical simulation  crash, and limit the propagation time to  essentially the early times of bifurcation.   This issue is discussed  below and at length in Ref. \cite{stetzler2025}. 

\begin{figure}
    \centering
    \begin{tabular}{ll}
         a)\hspace{1in} $V_r(q,t)$ & b) \hspace{1in}Ehrenfest \\
         \includegraphics[width=0.45\linewidth]{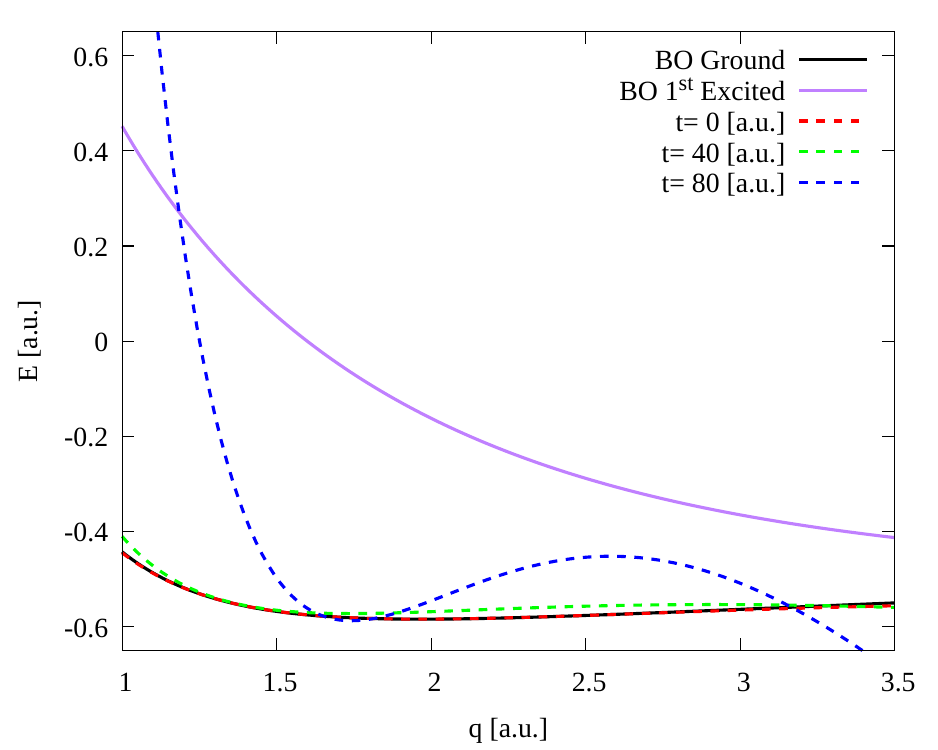} & \includegraphics[width=0.45\linewidth]{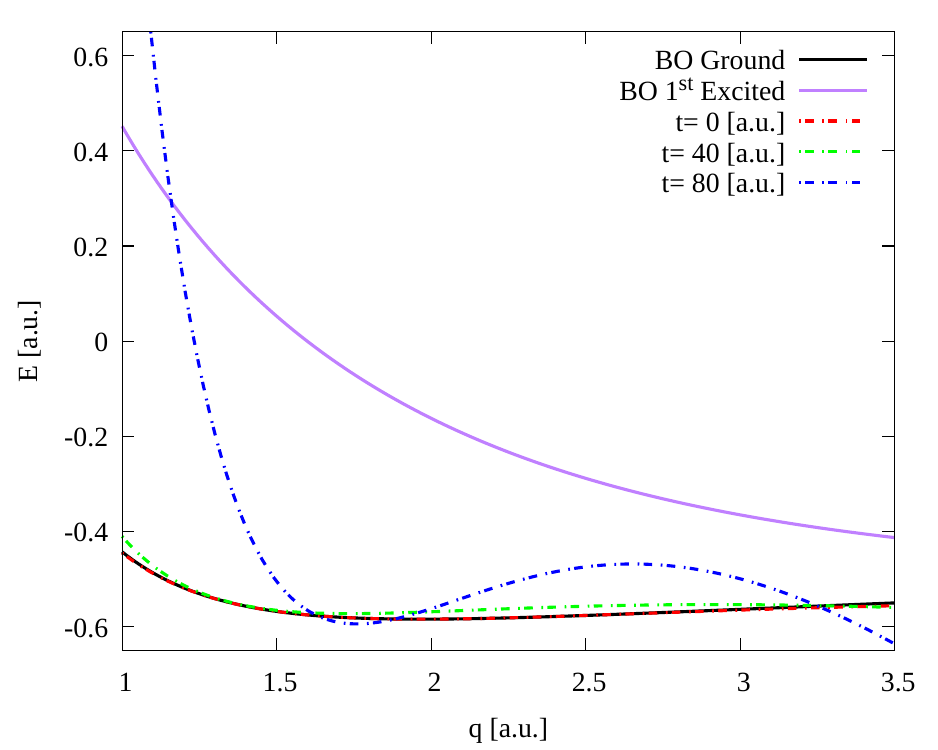}
    \end{tabular}
    \caption{Snapshots of (a)  $V_r$ (dashed lines) and (b) the Ehrenfest surfaces  (dot dashed lines) projected into the basis. In both panels the solid lines are the ground (black) and excited (purple) BO surfaces. Note the appearance of a "hill" centered around $q=2.5$ in both panels.}
    \label{fig:TDPES}
\end{figure}

\begin{figure}
    \centering
    \includegraphics[width=0.5\linewidth]{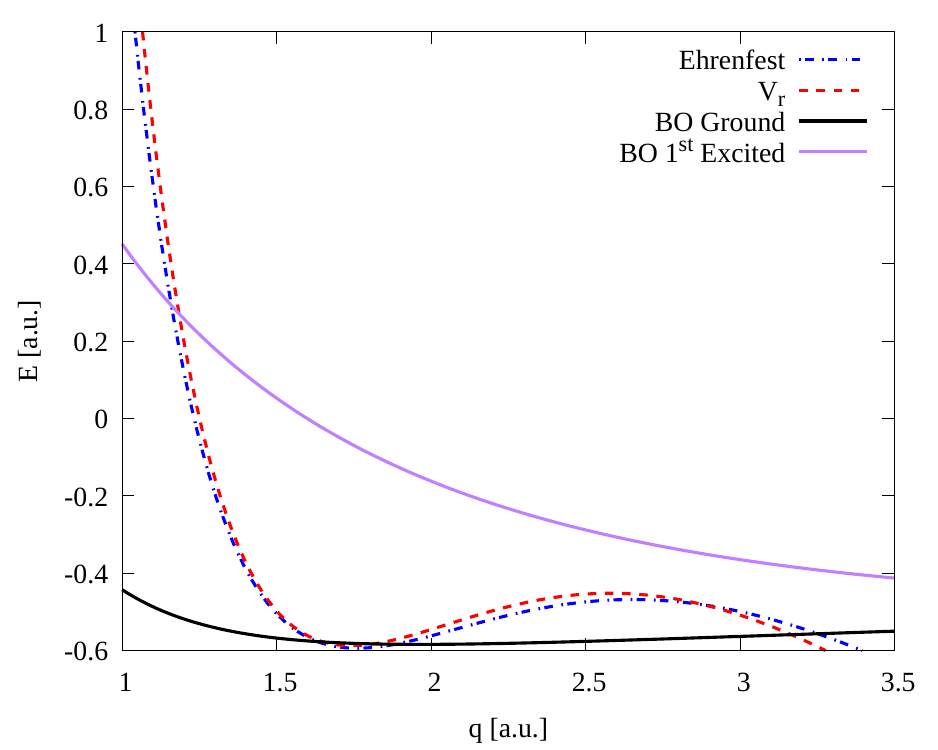}
    \caption{The Ehrenfest surface (dot dash) and  $V_r$ (dashed) at $t=80$ a.u.  The BO  surfaces are given for reference.}
    \label{fig:PESComparison}
\end{figure}

\begin{figure}
    \begin{tabular}{ll}
      a)\hspace{1in}   t=60 a.u. & b) \hspace{1in} t=80 a.u.  \\
         \includegraphics[width=0.47\linewidth]{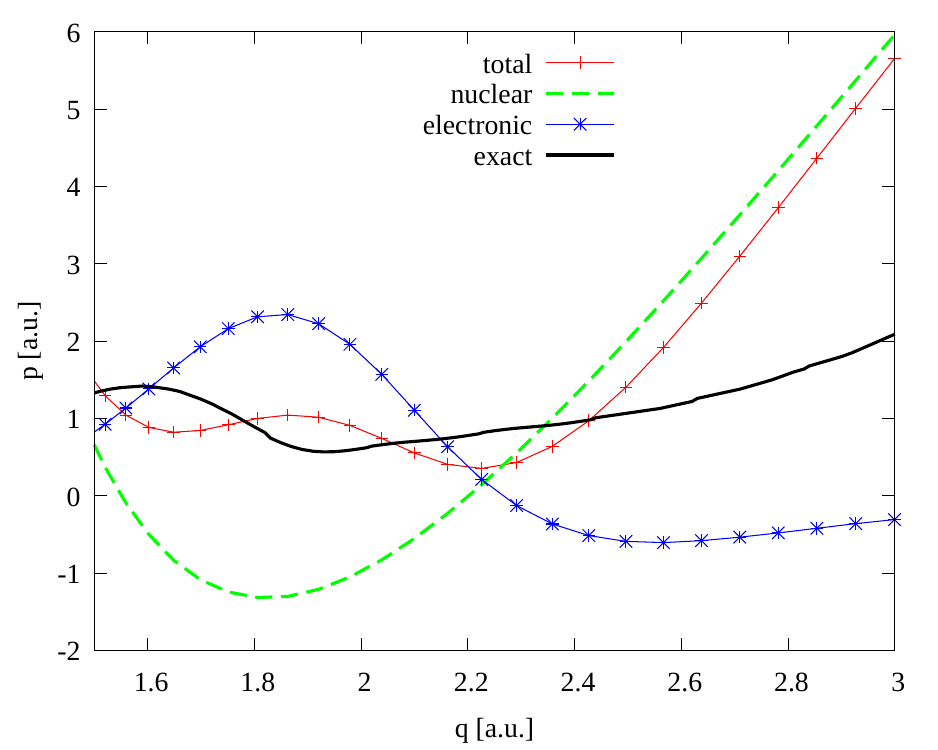} & \includegraphics[width=0.47\linewidth]{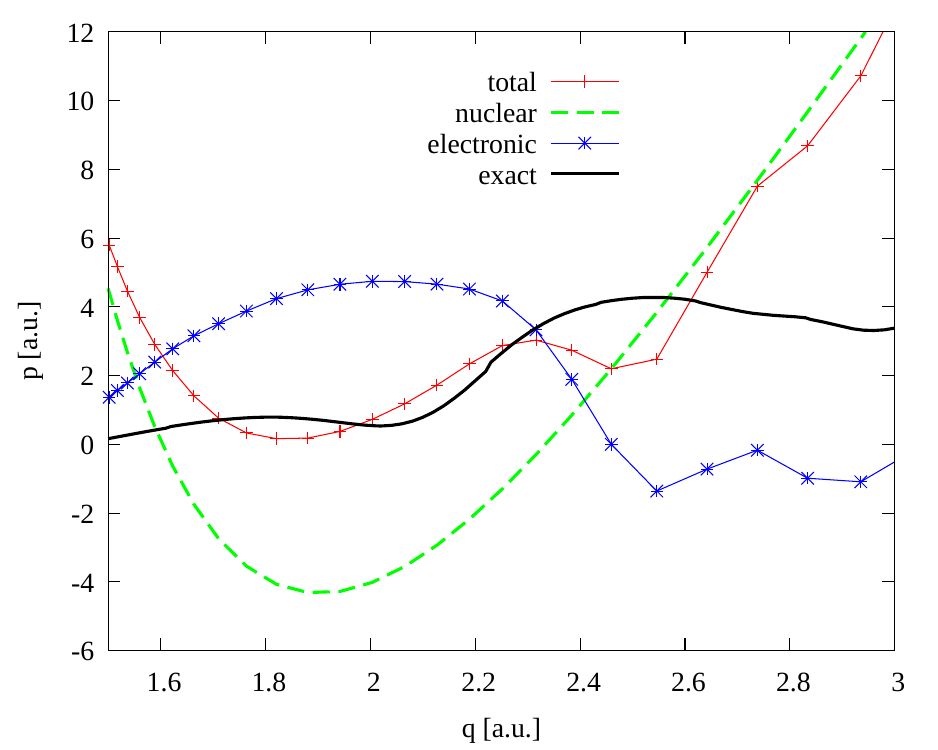}
    \end{tabular}
    \caption{The nuclear momentum components as functions of  $q$ for (a) $t=60$ a.u. and (b) $t=80$ a.u. In both panels: the black line represents the exact total nuclear momentum; the red curve is the total nuclear momentum in FENDy, the green dash is $p_\psi$ and the blue curve is $\overline{p_\Phi}$, 
      with  the dots marking the positions of the QTs. 
      }
    \label{fig:Momq}
\end{figure}

Now let us turn to the  'unique' feature of FENDy, i.e. the effective  potential, $V_d(q,t)$,  driving the dynamics of the nuclear wavefunction, defined here  as the expectation value of the electronic hamiltonian and coupling terms, Eq. (\ref{eq:He}).  
As a reminder, $V_d$, is a time-dependent complex function of the nuclear coordinate, with its real part, $V_r$, acting as the time-dependent PES on which $\psi(q,t)$ evolves, and its imaginary part, $V_i$, enforcing the normalization of the electronic wavefunction $\Phi(\bm R,q,t)$ for each $q$. The 'ideal' $V_r$  means that the residual nuclear momentum, $\overline{p_\phi}$ of Eq. (\ref{eq:p_phi_av}), in the electronic wavefunction and $V_i$ are both equal to zero.  We verified that in our model (one nuclear degree of freedom)   as implemented,   $V_r$  is, indeed,  'ideal'  in the {\it absence} of the laser pulse:   since the H$_2^+$  system is characterized by a large energy gap between the ground and excited electronic states, $V_r$ is essentially the same as the ground BO PES, $V_i=0$ and $\overline{p_\psi}=0$.     
However, with the laser pulse present,   $V_r$  exhibits features  responsible for the dissociation of the nuclear  wavefunction  due to the electron-nuclear coupling as illustrated in Fig. \ref{fig:TDPES}(a).  The BO surfaces are plotted alongside $V_r$ snapshots  for reference.  Given our "mean-field" like definition of $V_r$,   we  also compared  it  to the Ehrenfest surfaces  defined as,  $$V^{Erf}=\bra \Phi | \hat{H}^{BO}+\hat{V}^{ext} | \Phi \ket_{\bm{R}},$$ at each trajectory and projected into the same basis as $V_r$ (Fig. \ref{fig:TDPES}(b)).   The difference between  the two types of surface is in the electron-nuclear coupling due to the operators $\hat{D}^{(1)}$ and $\hat{D}^{(2)}$. 
For both types of surfaces in Fig. \ref{fig:TDPES}, we note the appearance of a "hill" in the region of  $q=[2.0,2.5]$ a.u. as the time  progresses. This "hill" corresponds to the position of the bulk of electronic excitation within the nuclear subspace; the force from this surface would give nuclear trajectories  at  $q<2$ a.u. the  negative momentum as seen in Fig. \ref{fig:Momq}. We find  $V_r$ to be  nearly identical to the Ehrenfest surfaces at early times with  the difference appearing at $t=80$ a.u. These snapshots are plotted together in Fig. \ref{fig:PESComparison}. 
The difference between the two is due to the larger electron-nuclear coupling at later times (also observed in the large $\bra \overline{p_\phi}\ket_q$ in Fig. \ref{fig:UWPops}(b)), as the Ehrenfest surface neglects these terms. Despite its simplicity, the  definition of $V_r$  implemented in this work is clearly 'sub-optimal'  as seen  in the large $\overline{p_\phi}$ it generates.   This conclusion  is consistent with  our  results for the   vibrationally nonadiabatic model   \cite{Fendy}.   
Alternative $V_r$, minimizing $\overline{p_\phi}$,  are  considered  in Ref. \cite{Fendy},  and   are currently under development in our group.   

\begin{figure}
    \centering
    \begin{tabular}{l l}
         a)  & b)  \\
         \includegraphics[width=0.47\linewidth]{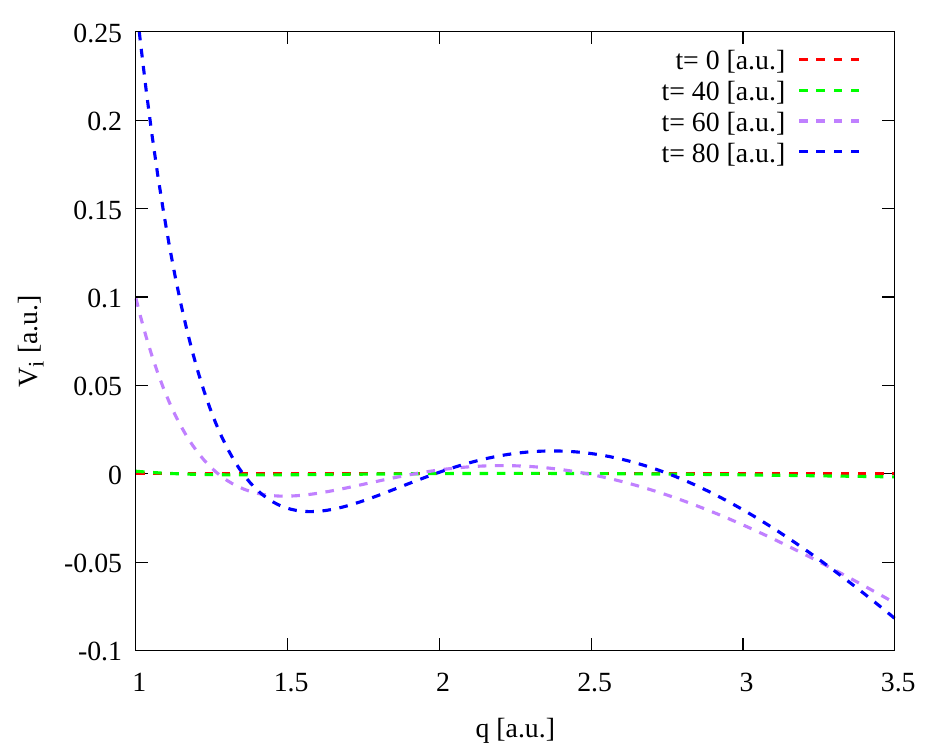} & 
\includegraphics[width=0.47\linewidth]{Figs/Momentum.pdf}
    \end{tabular}
    \caption{(a) Snapshots of the imaginary part of the time-dependent potential as a function of $q$. (b) The nuclear momentum components averaged over the nuclear sub-space as functions of time (same as Fig. (\ref{fig:UWPops}(b)).   
    }
    \label{fig:ImTDPES}
\end{figure}

Now let us examine the computed  $V_i(q,t)$  and  $\overline{p_\phi}(q,t)$ in more detail.  In our simulation $V_i(q,t)$, shown in Fig. \ref{fig:ImTDPES}(a),  is   very close to zero until the onset of the  wavefunction bifurcation  in the nuclear coordinate. This behavior is manifested in the  dynamics of  the nuclear momentum components, $p_\psi$, and $\bra \overline{p_\phi}\ket_q$ (Fig. \ref{fig:ImTDPES}(b)), with the former decreasing and latter increasing  after $t>40$ a.u. 
The nuclear momentum components as functions of $q$  are shown in 
Fig. \ref{fig:Momq} at  $t=60$ and $t=80$ a.u.  Note that in both instances  near the equilibrium bond length of $q=2$ a.u., which is the most populated region of the nuclear subspace,  $p_\psi$ is negative and  $\overline{p_\Phi}$ is positive; they add up to  the  total nuclear momentum that is close to the split-operator result.   Despite the discrepancies  between  the exact and FENDy  nuclear momenta  in Fig. \ref{fig:Momq},    their averages  are in reasonable agreement (Fig. \ref{fig:UWPops}(b))  given the nuclear density distribution.   If the laser is turned off, then  the computed  $\overline{p_\Phi}$ remains zero (not shown),   suggesting that in this model the residual nuclear momentum of the electronic wavefunction is introduced by the laser. Large values or  $\overline{p_\Phi}$  correspond to a large imaginary potential which is consistent with Eq. (\ref{eq:exactVi}). Without the laser pulse, the imaginary potential remains zero.

\begin{figure}
    \centering
    a) \includegraphics[width=0.45\linewidth]{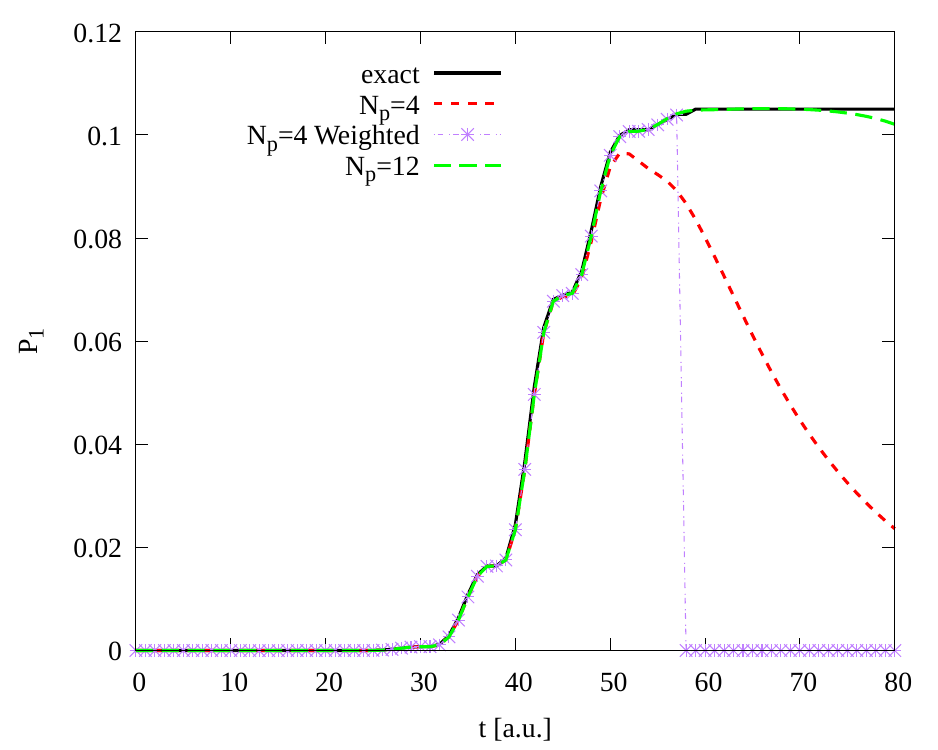} b) \includegraphics[width=0.45\linewidth]{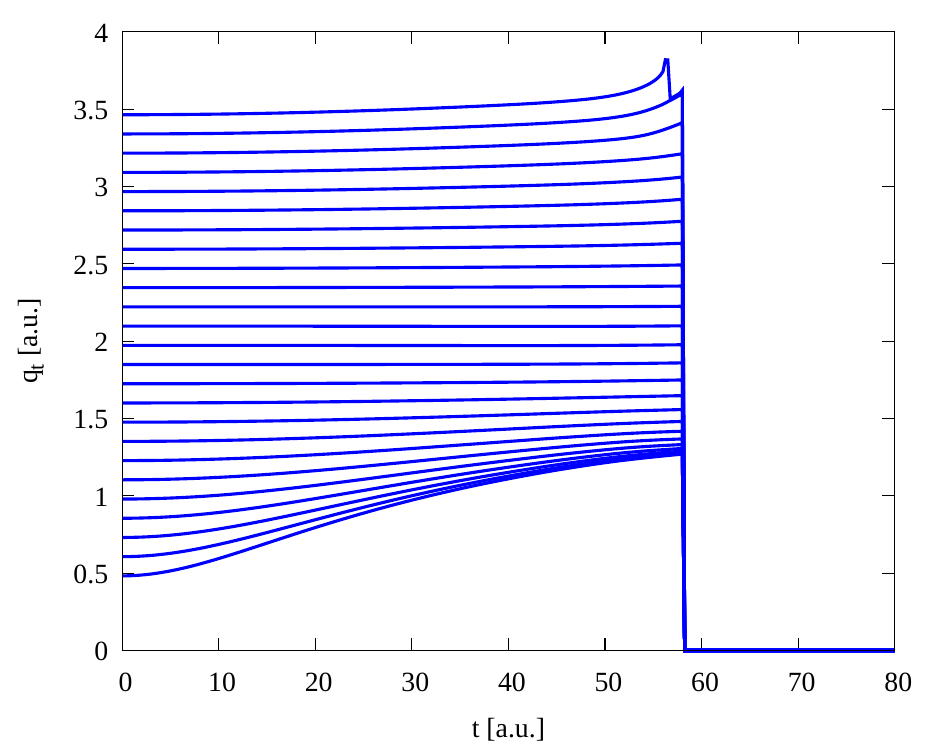}
    \caption{Panel a: The same as Fig. \ref{fig:UWPops}a, except the purple line is computed using weighted fits of $\mathbf{C}$. Panel b: The same as Fig. \ref{fig:traj}a but computed with weighted fits. The simulation crashed around $t=60$ a.u. due to accumulated noise in the electronic boundaries from the weighted fits.}
    \label{fig:WPops}
\end{figure}

Next we will discuss the challenges of stable propagation of the electronic coefficients $\mathbf{C}$. One already mentioned challenge is the unphysical 'erratic' values of $\mathbf{C}$ due to crossing of trajectories:  the approximate quantum force does not fully balance the forces acting on the trajectories on the steep potential wall.  This results in the interdispersed reflected and incoming subgroups of trajectories, each with different values of $\mathbf{C}$ and $p_\psi$, and essentially random values of derivatives of these quantities.  This artifact is mitigated with the low pass filter procedure described in Section \ref{sec:Implementation}. Another significant challenge of numerical XF are the non-trivial electronic boundary conditions in $q$ as consequence of the partial normalization condition, which must be maintained even in the regions of very low overall probability density. We demonstrate this by comparing the unweighted and weighted (with the nuclear density) fits of $\mathbf{C}$ (described in Section \ref{sec:BasisProjections}). In Fig. \ref{fig:WPops}(a), we present populations obtained with unweighted and weighted projections. As expected the weighted projections provide more accurate description of the nuclear-averaged behavior with a smaller Fourier basis,  but instabilities in the tails of the density eventually crash the simulation. These instabilities are more apparent in  Fig. \ref{fig:WPops}(b) where the tail trajectories behave erratically as a result of significant errors in $\overline{p_\phi}$.  In principle, this can be resolved using analytic boundary conditions,  which is non-trivial in the delocalized laser field. We mitigate the problem by fitting electronic coefficients independent of the nuclear density (i.e. 'unweighted fits'), improving the accuracy at the boundary regions at the cost of reduced accuracy in expectation values.

\begin{figure}
    \centering
    \includegraphics[width=0.6\linewidth]{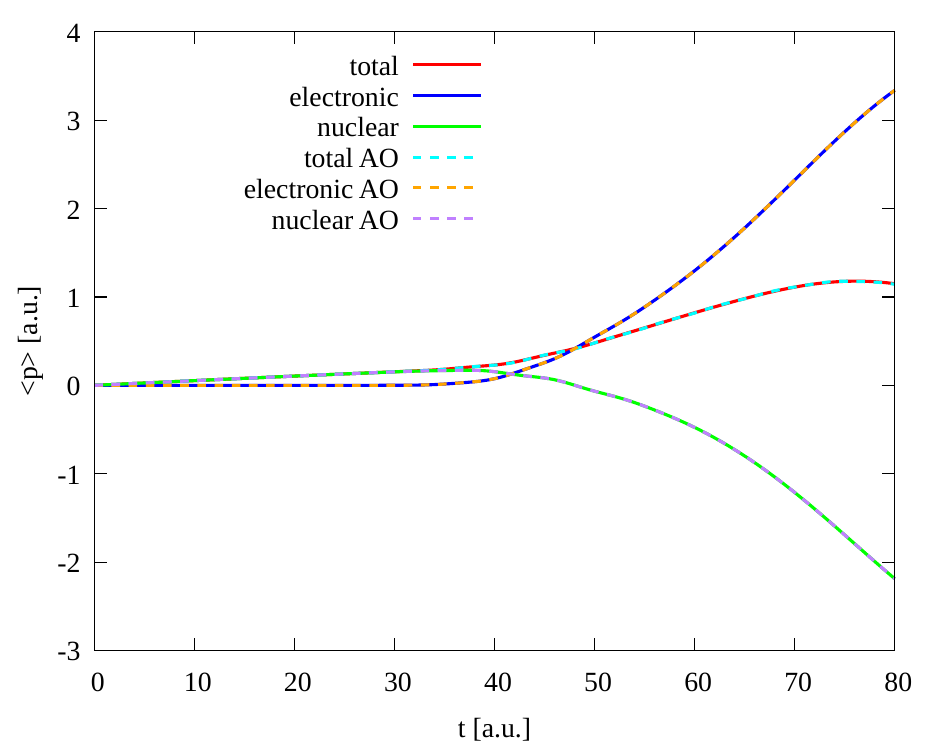}
    \caption{
    The nuclear momentum components computed in the eigenbasis and AO basis with $N_p$=8. The eigenbasis/AO basis  results are  shown as  with solid lines/dashes, respectively.} 
    \label{fig:AvMOM}
\end{figure}

Finally, we illustrate the equivalence of the atomic basis and eigenbasis. Performing  the dynamics directly in the atomic basis, without computing the eigenstates,  is one of the attractive features of the XF electron-nuclear dynamics.  
The average nuclear momenta  obtained in two FENDy implementations are displayed in Fig. \ref{fig:AvMOM}. We see that not just the total nuclear momenta computed in the eigenbasis and directly in the atomic basis match, but $p_\psi$ and $\bra\overline{p_\phi}\ket$ are identical as well. 
The comparison of  the orbital coefficients (Table \ref{tab:AOComp}) reinforces this point. 
The same simulation is performed using the AO representation and  we compare the two electronic wavefunctions obtained at $t=21$ a.u. at the central (25$^{th}$) trajectory  by projecting the eigenbasis wavefunction back into the AO basis.
As expected the AO coefficients show excellent agreement (better than $10^{-4}$).  

\setlength{\tabcolsep}{10pt}
\renewcommand{\arraystretch}{1}
\begin{table}[]
    \centering
    \begin{tabular}{c|c|c}
     AO coefficient & AO simulation & eigenbasis simulation \\\hline
     H$^L_{1s}$ & $-0.38251	-0.00063\imath$  & $-0.38259	-0.00052\imath$ \\
     H$^L_{2s}$ & $-0.25794	-0.00115\imath$ & $-0.25784	-0.00111\imath$ \\
     H$^R_{1s}$ & $-0.38168	+0.00040\imath$ & $-0.38169	+0.00052\imath$ \\
     H$^R_{2s}$ & $-0.25588	+0.00108\imath$ & $-0.25590	+0.00111\imath$ \\
     \end{tabular}
    \caption{Coefficients of 1s and 2s functions of  6-31G basis set on each atom (L, R)  on the central  trajectory at $t=21$ a.u.}
    \label{tab:AOComp}
\end{table}
\setlength{\tabcolsep}{6pt} 
\renewcommand{\arraystretch}{1}

\section{Conclusions} \label{sec:summary}  

In this paper, we presented a successful numerical implementation of FENDy in a molecular model with real Coulomb interactions and an external field. We demonstrate that the FENDy code is able to reproduce the exact dynamics obtained in a two-state model using the split-operator method. Most importantly, the method is developed without referencing the electronic eigenstates, and we demonstrated that the same result is achieved without solving the electronic structure. 
The biggest challenge of implementing the numerical XF was identified  as  the non-local effects associated with the electronic coefficients ${\mathbf C}$. Beyond the challenges of working with an unstructured grid in the trajectory representation of the nuclear wavefunction -- 
 something specific to the nuclear dynamics methods  employed  in this work --  we find the XF formalism itself presents significant numerical challenges when describing ${\mathbf C}$. 
 
 Firstly,  the diverging nuclear wavepackets in the H$_2^{+}$ model (and many other photochemical systems) lead to singularities in ${\mathbf C}$ due to the XF definition of the nuclear wavefunction, further explored in Ref. \cite{stetzler2025}. Secondly, the partial normalization condition leads to non-trivial electronic boundary conditions. In this work  we showed that the improper treatment of the electronic boundary (where the nuclear density is small) results in noise that propagates into the high density regions and crashes the calculation, and  that the use of unweighted fitting procedure  for  ${\mathbf C}$  alleviates this problem.  The  diverging nuclear dynamics may be less of a problem for larger systems in the regime of   electronic wavepackets, which is the envisioned regime of FENDy,  rather than a superposition of effectively only two states.  We also note that the use of basis set projections for $\grad_q \psi / \psi$, as in Ref. \cite{LQF}, eliminate this commonly cited (e.g. Refs \cite{neepanumericalxf,lorinxf}) source of numerical difficulty for XF. 
 
 Specific for FENDy with complex effective nuclear potential $V_d$,   we found  the $V_d$ derived for exact $\Phi$ led to errors in the finite basis representation. Thus we employed a simple definition of $V_d$, which is consistent with the finite basis representation and maintains the partial normalization condition. However, this definition results is not ideal as it leads to a large residual nuclear momentum of the electronic wavefunction. Better definitions of $V_r$, as well as the problem of diverging nuclear wavepackets and  improved description of the electronic boundaries  are the focus of the current and future investigation in our group.

\section*{Acknowledgement}
This material is based upon work supported by  the National Science Foundation of U.S.A. under Grant 
No.   CHE-2308922. 
\section*{Additional information}
{\bf Conflict of interests:}
The authors declare that they have no known competing financial interests or personal relationships that could have appeared to influence the work reported in this paper

{\bf Data availability: } The code described in this  work and used to generate the data is freely available on GitHub at:\url{https://github.com/julianps1/Factorized-electron-nuclear-dynamics-FENDy-/tree/main}

\appendix
\section{Interpolation Scheme} \label{sec:Interp}

In principle the code is compatible with on-the-fly electronic structure calculations; however, for computational efficiency we use a four-point Lagrange interpolation \cite{stegun1964} for electronic matrix elements.
In the AO representation, this is done for ${\mathbf H}^{BO}$ matrix elements. In the eigenbasis representation, this is done for all electronic matrix elements (except the overlap) as we do not have analytic expressions for the electronic eigenfunctions in $q$. 

We find approximate analytic forms of the boundary $q\rightarrow\infty$ to avoid computing a large number of interpolation points for dissociating trajectories, by assuming the wavefunction  approaching the non-interacting atomic limit at large $q>q_{cut}$. The cutoff, $q_{cut}$, is determined by the deviation of overlap matrix of the AOs below the preset tolerance, $\epsilon$, Table \ref{tab:num_params}. The energy and expansion coefficients are then: 
\be F(q):=Ae^{-\alpha q}+F_{atomic} \mbox{    if }q>q_{cut}, \ee
here $F_{atomic}$ is the value of the energy or expansion coefficient of the given state in the atomic limit for the eigenbasis. In the AO basis, these are the elements of ${\mathbf H}^{BO}$ in the atomic limit and include an additional $-1/q$ term scaled by the overlap of the basis functions (computed in the atomic limit) for the given matrix element. The exponential function is used to smoothly connect the asymptotic value to the numeric result at the cutoff. The scaling parameter $\alpha$ is computed from a 3-point finite difference at the cutoff. For AOs this is computed from the largest overlap matrix element and for the eigenbasis it is computed from the energy.
The parameter $A$ is chosen such that the analytic value is equal to the numeric value at $q_{cut}$ after obtaining $\alpha$.

In the eigenbasis representation we define analytic boundaries for the $q=0$ limit as the solutions become singular at $q=0$. The cutoff is chosen as $q_{cut}=0.5$ a.u. which gives reasonable accuracy in this model before the AOs become too degenerate. Here, the analytic form of the energy is chosen as
\be E(q):=aq^2+E_{0} \ee
where the quadratic parameter is defined such that the numeric and analytic solutions are equal at the cutoff, and $E_{0}$ is the energy at $q=0$. For the coefficients we add a linear parameter:
\be C(q):=aq^2+bq+C_{0}(q) \ee
where $C_0$ is the coefficient at $q=0$. The linear and quadratic terms are determined such that the function value and first derivative are equal at the cutoff. 

The analytic solutions at $q=0$ are obtained by a non-unitary transformation to the degenerate pairs of basis functions $\phi$: 
\bea 
g_{\lambda}:=\frac{1}{2}\phi_{\lambda}^{(1)} & + & \frac{1}{2}\phi_{\lambda}^{(2)} \nonumber, \\
u_{\lambda}:=\frac{1}{\sqrt{2\sigma}q}\phi^{(1)}_{\lambda} & - & \frac{1}{\sqrt{2\sigma}q}\phi^{(2)}_{\lambda},
\eea
where $s_\lambda$ is a quadratic coefficient obtained from,
\be \langle\phi_{\lambda}^{(1)}|\phi_{\lambda}^{(2)}\rangle_{\bm{R}}=1-s_\lambda q^2, \ee
here the numeric superscript indexes the atom with greek indexing the basis function. With this transformation, the new overlap matrix is non-degenerate as the new basis functions are orthogonal by construction and the singularities are in the analytic transformation matrix.  The analytic ${\mathbf D}^{(1)}$ and ${\mathbf D}^{(2)}$ matrix elements are obtained with a second order Taylor expansion at the cutoff to ensure they are smooth.

\bibliography{bib/JS1}

\end{document}